\PassOptionsToPackage{dvipsnames}{xcolor}
\documentclass[reprint,twocolumn,amsmath,amssymb,aps,floatfix]{revtex4-2}
\usepackage{graphicx}
\usepackage{dcolumn}
\usepackage{bm}
\usepackage{hyperref}
\usepackage{lipsum}
\usepackage{subcaption}  
\usepackage{caption}         
\usepackage{float}
\usepackage{tikz}
\usepackage{ragged2e}
\usepackage{placeins}
\usetikzlibrary{positioning, arrows.meta}  
\usepackage{relsize}  
\usepackage{overpic}  
\usepackage{placeins}
\usepackage{amssymb}
\usepackage{amsmath}
\usepackage{algorithm}
\usepackage{algpseudocode}
\usepackage{makecell}
\usepackage{bm}
\usepackage{placeins}
\usepackage[export]{adjustbox}
\usepackage{amsthm}

\theoremstyle{remark}

\usepackage{mathtools}
\usepackage{float}
\usepackage{tikz}
\usepackage{ragged2e}
\usetikzlibrary{positioning, arrows.meta}  
\usepackage{relsize}  
\usepackage{overpic}

\usepackage{tabularx}
\usepackage{booktabs}
\usepackage{array}
\usepackage{enumitem}
\usepackage{ragged2e}
\usepackage{makecell}
\usepackage{rotating}
\usepackage{xcolor}
\usepackage{array}
\usepackage{multirow}
\usepackage{graphicx}
\usepackage{relsize}
\newcommand{\simplot}[2][3.5cm]{%
  \parbox[c]{#1}{\centering
    \vspace*{3pt} 
    \includegraphics[width=#1]{#2}
    \vspace*{2pt}
  }%
}

\makeatletter
\let\CombinedSuppSection\section
\let\CombinedSuppSubsection\subsection
\let\CombinedSuppSubsubsection\subsubsection
\makeatother
\makeatletter

\renewcommand\section{%
  \@startsection{section}{1}{0pt}%
  {2.0ex plus 1ex minus .2ex}%
  {1.0ex plus .2ex}%
  {\normalfont\normalsize\bfseries\raggedright}%
}

\renewcommand\subsection{%
  \@startsection{subsection}{2}{0pt}%
  {1.5ex plus .8ex minus .2ex}%
  {0.8ex plus .2ex}%
  {\normalfont\normalsize\bfseries\raggedright}%
}

\makeatother
\usepackage{comment,textcase}
\usepackage[normalem]{ulem}
\newif\ifSupplementReferences
\makeatletter
\newcommand{\SupplementCite}[1]{%
  \begingroup
  \@fileswfalse
  \cite{#1}%
  \endgroup
}
\makeatother

\makeatletter
\let\CombinedTitle\title
\let\CombinedAuthor\author
\let\CombinedAffiliation\affiliation
\let\CombinedThanks\thanks
\let\CombinedEmail\email
\newcommand{\CombinedBeginSupplement}{%
  \clearpage
  \onecolumngrid
  \@booleanfalse\twocolumn@sw
  \def\title@column##1{##1}%
  \renewenvironment{widetext}{\par}{\par}%
  \let\section\CombinedSuppSection
  \let\subsection\CombinedSuppSubsection
  \let\subsubsection\CombinedSuppSubsubsection
  \setcounter{section}{0}\setcounter{subsection}{0}
  \setcounter{figure}{0}\setcounter{table}{0}\setcounter{equation}{0}
  \setcounter{footnote}{0}
  \renewcommand{\thefigure}{S\arabic{figure}}
  \renewcommand{\thetable}{S\arabic{table}}
  \renewcommand{\theequation}{S\arabic{equation}}
  \renewcommand{\theHsection}{supp.\arabic{section}}
  \renewcommand{\theHsubsection}{supp.\arabic{section}.\arabic{subsection}}
  \renewcommand{\theHequation}{supp.\arabic{equation}}
  \renewcommand{\theHfigure}{supp.\arabic{figure}}
  \renewcommand{\theHtable}{supp.\arabic{table}}
  \theoremstyle{plain}
  \setcounter{lemma}{0}
  \let\title\CombinedTitle
  \let\author\CombinedAuthor
  \let\affiliation\CombinedAffiliation
  \let\thanks\CombinedThanks
  \let\email\CombinedEmail
  \let\maketitle\frontmatter@maketitle
  \frontmatter@init
  \global\setbox\absbox\box\voidb@x
}
\makeatother

\begin{document}

\makeatletter
\let\auto@bib@innerbib\@empty
\let\auto@bib\@empty
\makeatother

\title{Phase-Sum Coupling for Gradient-Flow Dynamics in Oscillator Ising Machines}

\author{Abir~Hasan\textsuperscript{1\#},
Sajib~Bain\textsuperscript{1\#},
E.M.~Hasantha~Ekanayake\textsuperscript{1}, Kyle~Lee\textsuperscript{2}, Kerem~Camsari\textsuperscript{2},
Nikhil~Shukla\textsuperscript{1*}}

\affiliation{%
\textnormal{\textsuperscript{1}}University of Virginia, Charlottesville, VA, USA\\
\textnormal{\textsuperscript{2}}\mbox{University~of~California,~Santa~Barbara,~Santa~Barbara,~CA,~USA}\\
\textnormal{\textsuperscript{*}Email: ns6pf@virginia.edu}\\
\textnormal{\textsuperscript{\#}Equal Contribution}
}

\begin{abstract}
Oscillator Ising machines are commonly viewed as dynamical systems that minimize an energy function, yet this interpretation is not generally valid for nonlinear oscillator hardware. Here, we show that conventional phase-difference coupling requires a stringent harmonic-by-harmonic quadrature condition between the oscillator waveform and its phase response. In contrast, symmetric phase-sum interactions intrinsically admit gradient-flow dynamics without this constraint. When combined with second-harmonic injection, the canonical sine phase-sum interaction recovers the Ising Hamiltonian on the binary phase manifold while defining a distinct continuous energy landscape. However, the physical realization of phase-sum coupling can introduce unwanted cross-harmonic interactions that compromise the gradient-flow structure. We derive a Nyquist-like phase-domain sampling criterion that specifies the number of coupling paths required to eliminate these interactions within a prescribed harmonic bandwidth. Guided by this result, we develop a CMOS architecture combining phase-resolved oscillator taps with second-harmonic-modulated coupling. Transistor-level simulations demonstrate an approximately fifteenfold reduction in non-gradient curl compared with conventional phase-difference coupling, while simulations of oscillator networks validate the resulting phase-sum dynamics. These results establish interaction symmetry as an additional design degree of freedom for engineering gradient-flow dynamics in nonlinear oscillator Ising machines.

\end{abstract}                             
\maketitle
\twocolumngrid 

\section*{Introduction}
\noindent The prospect of leveraging the natural dynamics of coupled oscillatory networks to solve challenging combinatorial optimization problems has driven the development of oscillator Ising machines (OIMs) \cite{wang2019oim,mohseni2022ising,vadlamani2020physics, todri2024computing,csaba2020coupled}. A wide range of physical oscillator platforms have been proposed for this purpose, including harmonic oscillators, on which the theoretical framework was originally established, optical parametric oscillators~\cite{mcmahon2016fully,inagaki2016coherent}, spinwave and surface-acoustic-wave Ising machines~\cite{litvinenko2023spinwave,litvinenko202550}, electrical LC oscillators~\cite{chou2019analog}, and phase-transition nano-oscillators~\cite{dutta2021ising}, as well as more hardware-friendly implementations such as ring oscillators~\cite{moy20221,graber2024integrated,lo2023ising,cilasun2025coupled}. 

In the canonical formulation based on the Kuramoto model for harmonic oscillators \cite{wang2019oim}, the phase dynamics take the form
\begin{equation}
    \dot{\theta}_i = -K\sum_{j\neq i} W_{ij} \sin(\theta_i - \theta_j) - K_s \sin(2\theta_i),
    \label{eq:intro_oim1}
\end{equation}
where $\theta_i$ denotes the phase of the $i^{\mathrm{th}}$ oscillator, $W_{ij}$ encodes the coupling weights corresponding to the Ising spin interactions, and $K$ and $K_s$ denote the coupling and the second-harmonic injection (SHI) strengths, respectively. The pairwise interaction function is sinusoidal and hence odd. Under suitable forcing mechanisms such as SHI, such systems admit a gradient-flow interpretation, with the dynamics performing gradient descent on an effective energy landscape whose minima encode solutions to the Ising Hamiltonian. This energy-based interpretation underpins much of the theoretical framework used to understand and design OIMs, but implicitly assumes that the physical oscillator network preserves the gradient structure of the ideal phase model.

In this work, we first show that this gradient-flow interpretation is not generic for conventional oscillator coupling and can break down in a broad class of practical implementations. For phase-difference interactions of the form \(H(\theta_i-\theta_j)\), where, $H(\cdot)$ denotes a general interaction function derived from the phase-response formalism. It can be demonstrated that a \emph{necessary} condition for a gradient-flow representation requires the interaction function be odd, up to an additive constant. Within the phase-response framework, this requirement translates into a \emph{harmonic-by-harmonic quadrature relation between the infinitesimal phase response curve (iPRC) of the oscillator}~\cite{schultheiss2011phase} \emph{and the injected perturbation waveform}. This condition is generally not satisfied by nonlinear oscillators whose waveforms and phase responses contain multiple harmonics. We term the resulting violation \emph{harmonic misalignment}: corresponding harmonics deviate from quadrature, generating finite even components in the effective interaction function, and consequently, non-conservative phase dynamics. 


We then show that dynamical Ising machines (DIMs), recently proposed by the authors~\cite{cdc9-y234} and based on an alternative phase-sum interaction of the form \(H(\theta_i+\theta_j)\), circumvent the harmonic misalignment constraint. Specifically, the DIM dynamics can be expressed as,
\begin{equation}
    \dot{\theta}_i
    =
    -K\sum_{j\neq i}W_{ij}\sin(\theta_i+\theta_j)
    -
    K_s\sin(2\theta_i).
    \label{eq:intro_dim}
\end{equation}
where the phase-sum based interactions are complemented by a SHI.

Moreover, for symmetric coupling weights, phase sum interactions in DIMs directly admit a scalar pair potential for \emph{arbitrary integrable \(H\)}, and therefore generate gradient-flow dynamics without requiring the harmonic-by-harmonic quadrature condition imposed by phase-difference coupling. 
While the computational model for an Ising machine based on phase-sum interactions was proposed by the authors in prior work~\cite{cdc9-y234}, here, we develop a quadrature-resolved circuit architecture that realizes signed phase-sum interactions and validates the proposed mechanism using transistor-level CMOS simulations. Together, these results establish interaction symmetry as a design degree of freedom for engineering gradient-flow Ising dynamics in nonlinear oscillator hardware.


\section*{Results}
\subsection{Breakdown of Gradient Flow in Conventional OIMs}
\noindent \textit{Condition for Gradient-Flow Dynamics}---We first evaluate the gradient-flow dynamics in conventional OIMs by considering a network of weakly coupled oscillators described by the reduced phase dynamics,
\begin{equation}
\dot{\theta}_i = \omega_i + \sum_{j} W_{ij}\, H(\theta_j - \theta_i),
\label{eq:phase_dynamics}
\end{equation}
where, $\omega_i$ is the natural frequency of the oscillator $i$, $H(\cdot)$ is the interaction function obtained from the phase response formalism~\cite{ko2009phase} and $W_{ij}$ denotes the effective coupling coefficient. We assume symmetric coupling, $W_{ij}=W_{ji}$. For simplicity, we set $\omega_i=1$ for all $i$.

As discussed above, the traditional operation of OIMs relies on the dynamics admitting a gradient-flow representation, i.e., $\dot{\theta}_i = -\nabla_{\theta_i} E$ for some scalar energy function $E(\boldsymbol{\theta})$ whose fixed points~\cite{bashar2023stability, cheng2024control} encode the Ising configurations. Equivalently, the induced vector field must be curl-free (see Supplementary Note 1), which implies
\begin{equation}
\frac{\partial \dot{\theta}_i}{\partial \theta_j} = \frac{\partial \dot{\theta}_j}{\partial \theta_i}, \quad \forall i,j.
\label{eq:curl_condition}
\end{equation}
Taking the partial derivatives of the phase dynamics in Eq.~\eqref{eq:phase_dynamics}, we obtain
\begin{equation}
\frac{\partial \dot{\theta}_i}{\partial \theta_j}
=
W_{ij}\, H'(\theta_j - \theta_i),
\qquad
\frac{\partial \dot{\theta}_j}{\partial \theta_i}
=
W_{ji}\, H'(\theta_i - \theta_j).
\end{equation}
Under symmetric coupling, and in the absence of additional phase shifts from the coupling elements (i.e., for effectively resistive coupling), the condition in Eq.~\eqref{eq:curl_condition} reduces to
\begin{equation}
\begin{split}
H'(\Delta_{ji}) = H'(\Delta_{ij})
\Rightarrow\,\, H(\Delta_{ji}) = -H(\Delta_{ij}) +C,
\end{split}
\end{equation}

\begin{figure*}
    \centering
    \includegraphics[width=0.9\linewidth]{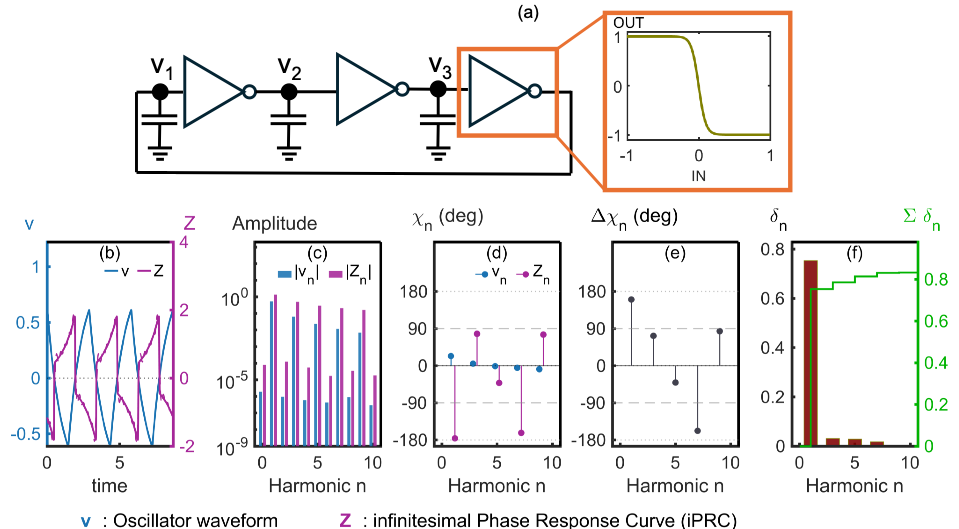}\\
    \caption{\justifying \textbf{Harmonic analysis of a three-stage (N=3) ring oscillator in a coupled two-oscillator system}. (a) Schematic of a three-stage ring oscillator. (b) Time-domain waveform $v(t)$ (left axis) and iPRC $Z(t)$ (right axis). (c) Amplitude; and (d) Phase spectra of the first ten harmonics of $v(t)$ and $Z(t)$, respectively. (e) Harmonic-wise phase difference, $\Delta\chi_n = \chi_n^Z - \chi_n^s$ (wrapped to $[-\pi,\pi]$), showing deviations from quadrature. (f) Normalized harmonic-resolved contributions $\delta_n$ to the even (non-gradient) component (left axis) and cumulative deviation $\delta$ (right axis).}
    \label{fig:Fourier Spectra}
\end{figure*}

\noindent where, $\Delta_{ij} = \theta_i - \theta_j$. Thus, a gradient-flow representation exists only if the interaction function is odd up to an additive constant. Such a constant, even if present, produces only a uniform phase drift and does not affect the integrability condition, which depends on $H'$. Any non-constant even component of $H$ induces a finite curl in the phase dynamics, precluding the resulting vector field from being written as the gradient of a scalar energy.\\

\noindent\textit{Harmonic Misalignment}---To characterize the symmetry properties of $H(\cdot)$, we note that, in the weak-coupling regime, the interaction function can be expressed via the phase response formalism~\cite{schultheiss2011phase,ko2009phase,bhansali2009gen} as
\begin{equation}
H(\Delta_{ij}) = \frac{1}{T} \int_0^T Z(t)\, s\!\left(t + \Delta_{ij}\right)\, dt,
\label{eq:H_def}
\end{equation}
where $Z(t)$ denotes the iPRC of the oscillator and $s(t)$ is the perturbation signal arising from coupling to other oscillators. Equation~\eqref{eq:H_def} shows that $H(\Delta_{ij})$ is given by the temporal overlap between the phase response and a phase-shifted version of the injected waveform.

Writing the iPRC and injected waveform in their amplitude--phase Fourier forms,
\begin{equation}
Z(t)=\sum_{n\geq 1}\alpha_n\cos(nt-\chi_n^Z),
\qquad
s(t)=\sum_{n\geq 1}\beta_n\cos(nt-\chi_n^s),
\end{equation}
where $\alpha_n$ and $\beta_n$ denote the harmonic amplitudes and
$\chi_n^Z$ and $\chi_n^s$ their respective phases, substitution into
Eq.~\eqref{eq:H_def} and averaging over one period gives
\begin{equation}
H(\Delta_{ij})
=
\sum_{n\geq1}
\frac{\alpha_n\beta_n}{2}
\cos\!\left[
n\Delta_{ij}+\chi_n^Z-\chi_n^s
\right].
\end{equation}
Expanding the cosine term,
\begin{align} 
\cos(n\Delta_{ij} &+ \Delta\chi_n)=  \notag \\
 &\cos(n\Delta_{ij})\cos(\Delta\chi_n) - \sin(n\Delta_{ij})\sin(\Delta\chi_n), \notag
\end{align}
where $\Delta\chi_n = \chi_n^Z - \chi_n^s$. The $\cos(n\Delta_{ij})$ terms are even in $\Delta_{ij}$ and therefore constitute the even component of $H(\Delta_{ij})$ (see Supplementary Note~2 for detailed derivation). Imposing the oddness condition $H(\Delta_{ji}) = -H(\Delta_{ij})$ requires that these even components vanish. Furthermore, since $\{\cos(n\Delta_{ij})\}$ are linearly independent over one period, the coefficient of each harmonic must vanish individually, yielding
\begin{equation}
\cos(\Delta\chi_n) =\cos(\chi_n^Z - \chi_n^s) = 0,
\qquad \forall\, n \text{ with } \alpha_n \beta_n \neq 0.
\label{eq:quadrature_condition}
\end{equation}
Equivalently,
\[
 \chi_n^Z - \chi_n^s = \pm \frac{\pi}{2} \; (\mathrm{mod}\,\pi),
\qquad \forall\, n \text{ with } \alpha_n \beta_n \neq 0.
\]
Thus, gradient-flow dynamics require \emph{harmonic-by-harmonic quadrature between the iPRC and the injected waveform}. Any violation of this condition, which we term \emph{harmonic misalignment}, generates a nonconstant even component in $H(\Delta_{ij})$, and hence, non-conservative phase dynamics. 
\begin{figure*}[t]
\centering
\resizebox{\textwidth}{!}{%
{\renewcommand{\arraystretch}{1.1}\setlength{\tabcolsep}{4pt}%
\scriptsize
\setlength{\arrayrulewidth}{1pt}%
\begin{tabular}{|c|w{c}{3.8cm}|w{c}{3.8cm}|w{c}{3.8cm}|w{c}{3.8cm}|}
\hline
& \rule{0pt}{30pt}\textbf{\normalsize \shortstack{Van der Pol\\Oscillator}}
& \textbf{\normalsize \shortstack{Duffing-Van der Pol\\Oscillator}}
& \textbf{\normalsize \shortstack{LC\\Oscillator}}
& \textbf{\large \textbf{\normalsize \shortstack{Threshold Switching\\ Relaxation Oscillator}}} \\[8pt]
\hline

\rule{0pt}{25pt} \large \rotatebox[origin=c]{90}{\textbf{Dynamics}}\rule{0pt}{25pt}
  & \parbox[c]{3.2cm}{\centering\vspace{4pt}%
      $\ddot{x} + \mu(x^{2}-1)\dot{x} + x = 0$%
      \vspace{4pt}}
  & \parbox[c]{3.2cm}{\centering\vspace{4pt}%
      $\ddot{x} + \mu(x^2-1)\dot{x} + bx + ax^{3} = 0$\\[2pt]
      \textit{($\mu=0.01,\;a=0.01$)}%
      \vspace{4pt}}
  & \parbox[c]{3.8cm}{\centering\vspace{4pt}%
      $\begin{array}{r@{{}={}}l}
        \dot{v} & \tanh(kv) - i - a v \\[1pt]
        \dot{i} & b v \\[2pt]
      \end{array}$\\[2pt]
      \textit{($a=1,\;b=1$)}%
      \vspace{4pt}}

& \parbox[c]{3.8cm}{\centering\vspace{4pt}%
    $\begin{array}{l}
      \dot{v} = \\[2pt]
      \left\{\begin{array}{@{}l@{\;}l@{}}
        (V_{dd}\!-\!v)g_m \!-\! g_s v, & {\scriptstyle v < v_{\mathrm{th}2}} \\[3pt]
        -g_s v,                          & {\scriptstyle v > v_{\mathrm{th}1}}
      \end{array}\right.
    \end{array}$\\[4pt]
    \textit{($g_s\!=\!0.01,\;V_{dd}\!=\!1$)}%
    \vspace{4pt}}  \\
 
\hline

\Large \rotatebox[origin=c]{90}{$\boldsymbol{\delta}$}
  & \simplot[3.5cm]{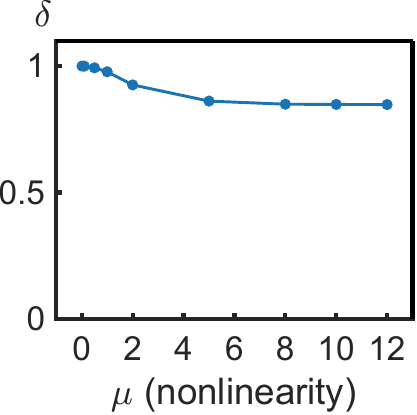}
  & \simplot[3.5cm]{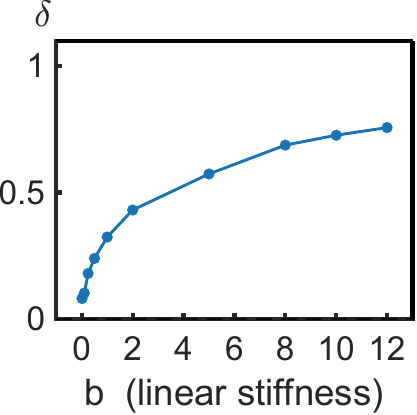}
  & \simplot[3.5cm]{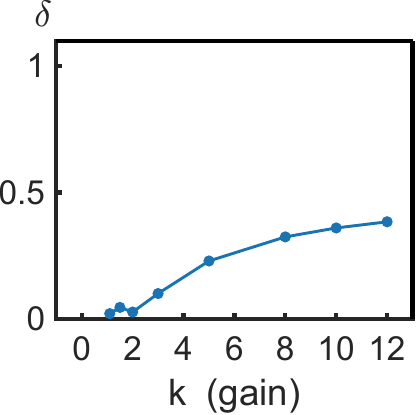}
  & \simplot[3.5cm]{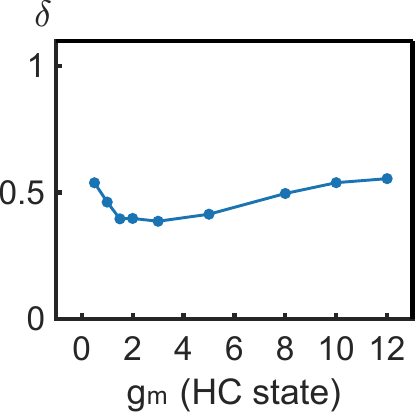} \\
\hline
\end{tabular}}}
\caption{\justifying \textbf{Deviation from gradient-flow in different oscillator topologies.} State-space equations and $\delta$ as a function of the oscillator-specific parameter for four oscillator topologies: Van der Pol ~\cite{barron2009synchronization}, Duffing-Van der Pol~\cite{chen2022van}, LC~\cite{chou2019analog}, and threshold switching relaxation oscillators~\cite{maffezzoni2015modeling}. Parameters: $\mu$ denotes the Van der Pol nonlinearity; $a$, $b$, and $\mu$ denote the cubic coefficient, stiffness, and nonlinear-damping in the Duffing-Van der Pol oscillator, respectively; $k$, $a$, and $b$ denote the LC gain, conductance, and inductance scaling parameters, respectively; and $g_m$, $g_s\,(=0.01,  \text{here})$, and $V_{dd}\,(=1, \text{here})$ denote the high-conductance (HC) state, discharge conductance, and supply voltage, respectively, for the threshold-switching relaxation oscillator. The peak-to-peak amplitude is set to $0.6$ by the switching thresholds $v_{th,1}=0.2$ and $v_{th,2}=0.8$.}
\label{tab:oscillators}\vspace{-15pt}
\end{figure*}\\

\noindent \textit{Coupled Pair of Ring Oscillators}---As a representative example of a non-linear oscillator platform with broad experimental relevance, we consider a network of ring oscillators (ROs), which have been widely explored as physical realizations of Ising spins~\cite{elmitwalli2025cmos,han2025digital,liu20241024,lo2023ising}. While these demonstrations highlight the practical utility and performance of RO-based Ising machine platforms, the question we address here focuses on whether the oscillator interactions admit an exact gradient-flow interpretation. A ring oscillator consists of an odd number of inverter stages ($N=2n+1$ where $n \in \mathbb{N}$), typically implemented using CMOS inverters and connected in a loop, giving rise to a stable limit cycle sustained by regenerative switching.

We consider a three-stage ring oscillator, illustrated in Fig.~\ref{fig:Fourier Spectra}(a), whose dynamics can be expressed using a set of coupled nonlinear differential equations determined by the inverter characteristics and inter-stage coupling (see Supplementary Note~3 for details of the ring oscillator dynamics). The output of each oscillator can be accessed at any inverter stage (referred to as a tap), with each tap providing a time-shifted version of the same periodic waveform. In a coupled network, signals from one oscillator may be injected into any of the available taps of another.

As discussed earlier, under weak coupling the phase dynamics reduce to pairwise interactions governed by the relative phase difference and the choice of injection and observation taps. In this regime, the interaction can be expressed as
\[
H_{r \leftarrow m}(\Delta_{ij}) = H_0(\Delta_{ij} + \beta_{rm}),
\]
where $H_0(\cdot)$ is a base interaction function determined by the intrinsic waveform and iPRC of the oscillator, and $\beta_{rm}$ denotes the phase offset associated with the selected tap configuration. Here, $r$ denotes the receiving tap associated with the iPRC, while $m$ denotes the transmitting tap from which the perturbation waveform is generated (see Supplementary Note~3). Thus, all coupling configurations correspond to phase-shifted versions of a single underlying interaction law.

We consider phase-preserving (i.e., resistive) coupling through the same taps at which the oscillator output is measured, corresponding to $\beta_{rm}=0$ (ferromagnetic (FM) coupling). Fig.~\ref{fig:Fourier Spectra}(b) shows the time-domain waveform and iPRC of a three-stage ring oscillator where each inverter has a gain of $k=70$, while Figs.~\ref{fig:Fourier Spectra}(c) and (d) present the corresponding amplitude and phase spectra for the first ten harmonics. Fig.~\ref{fig:Fourier Spectra}(e) plots the phase difference $\Delta\chi_n = \chi_n^Z - \chi_n^s$, revealing that multiple harmonics deviate from quadrature. These deviations induce finite even components in the interaction function, thereby violating the condition for gradient-flow dynamics established above.

To quantify the deviation from gradient dynamics, we introduce a normalized measure based on the even component of the interaction function (see Supplementary Note~2 for details of the metric formulation). Specifically, we define
\begin{equation}
\delta = \sum_{n \ge 1} \delta_{n} = \frac{\sum_n n \alpha_n \beta_n |\cos(\chi_n^Z - \chi_n^s)|}{\sum_m m \alpha_m \beta_m},
\label{eq:non_grad_metric}
\end{equation}
where, 
\[
\delta_n = \frac{ n \alpha_n \beta_n |\cos(\chi_n^Z - \chi_n^s)|}{\sum_{m \geq 1} m \alpha_m \beta_m},\qquad \delta_n,\,\delta \in [0,1]
\]

By construction, $\delta = 0$ for purely odd interaction functions, corresponding to ideal gradient-flow dynamics, while $\delta = 1$ indicates a purely even interaction and maximal deviation from energy-based behavior. Thus, $\delta$ directly quantifies the strength of non-gradient contributions induced by harmonic misalignment.

Fig.~\ref{fig:Fourier Spectra}(f) shows the normalized harmonic-resolved contributions $\delta_n$ and the cumulative $\delta$ for the coupled ring oscillators. The first harmonic already exhibits a substantial even component and dominates the overall deviation, indicating that non-gradient behavior is driven primarily by low-order harmonics. Furthermore, since changing the transmit or receive tap only shifts the interaction function in phase, the magnitude-based quantities $\delta_n$ and $\delta$ are invariant to the choice of taps. A corresponding analysis for a longer-chain ring oscillator ($N=25$) is provided in Supplementary Note~4, where similar behavior is observed.


To place these results in a broader context, we evaluate $\delta$ across four representative oscillator models---Van der Pol~\cite{barron2009synchronization}, Duffing--Van der Pol~\cite{chen2022van}, LC~\cite{chou2019analog}, and threshold switching relaxation oscillators~\cite{parihar2015synchronization,maffezzoni2015modeling,maher2024cmos,lee2024demonstration}---which are relevant as candidate oscillator platforms for OIMs. These models span a range of nonlinearities and interaction characteristics, enabling a comparative assessment of the propensity of their interaction dynamics to exhibit gradient-flow behavior. We note that the LC oscillator considered here corresponds to a practical realization in which a parallel active gain element compensates the resistive losses of the resonator~\cite{lai2004capturing}.

Fig.~\ref{tab:oscillators} summarizes the governing state-space equations and the corresponding evolution of $\delta$ for these oscillator models. LC oscillators at low gain exhibit small values of $\delta$, indicating interactions close to gradient flow. In contrast, the Van der Pol oscillator exhibits large $\delta$ across a wide range of nonlinear damping strengths. Duffing--Van der Pol oscillators show an increasing deviation from gradient flow with increasing linear stiffness. Threshold-switching relaxation oscillators also show appreciable non-gradient contributions, with $\delta > 0.3$ across the parameter range considered. These findings motivate our investigation of phase-sum interactions, which offer an alternative route to gradient-flow dynamics without imposing the stringent harmonic-by-harmonic quadrature constraint of conventional phase-difference coupling.

\subsection{Phase-Sum Interactions as an Alternative Route to Certifiable Gradient Flow}

\noindent \textit{Phase-Sum Interactions Eliminate the Harmonic Misalignment Constraint}---The preceding analysis shows that gradient flow breaks down when harmonic misalignment between the oscillator waveform and its phase response generates interaction components that violate the odd-symmetry requirement imposed by phase-difference coupling. This motivates a more fundamental question: \textit {can gradient flow be guaranteed in a nonlinear oscillator network without mandating harmonic-by-harmonic alignment?}

To circumvent the harmonic-by-harmonic constraint imposed by phase-difference coupling, we consider a symmetric pairwise phase-sum interaction,
\begin{equation}
\dot{\theta}_i^{(ij)}
=
\dot{\theta}_j^{(ij)}
=
-W_{ij}H(\theta_i+\theta_j),
\qquad W_{ij}=W_{ji}.
\label{eq:phase_sum_pair}
\end{equation}
For any integrable interaction function \(H\), Eq.~\eqref{eq:phase_sum_pair} is generated by the scalar pair potential
\begin{equation}
E_{ij}
=
W_{ij}
\int^{\theta_i+\theta_j} H(\Theta)\,d\Theta,
\label{eq:general_phase_sum_potential}
\end{equation}
since
\(
-\partial E_{ij}/\partial\theta_i
=
-\partial E_{ij}/\partial\theta_j
=
-W_{ij}H(\theta_i+\theta_j)
\).
Thus, unlike phase-difference coupling, phase-sum coupling admits gradient-flow dynamics without requiring \(H\) to be odd, and therefore without imposing harmonic-by-harmonic quadrature between the oscillator waveform and its phase response. We also note that the same applies to $H(n(\theta_i+\theta_j))$, where ($n\in \mathbb{Z}$) \\
\begin{figure}[!h]
    \centering
    \includegraphics[width=1\linewidth]{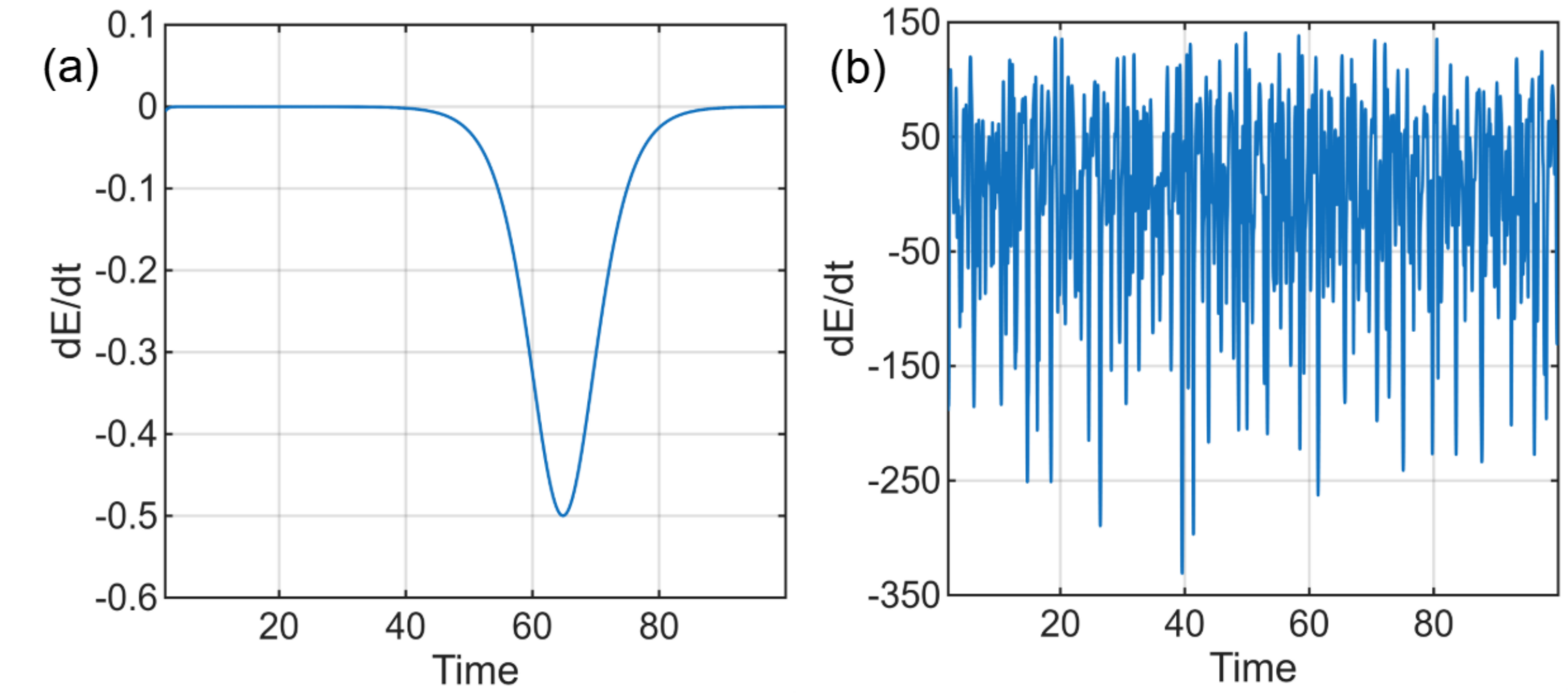}
    \caption{\justifying \small
    \textbf{Energy-rate comparison of phase-sum and phase-difference dynamics.} $\dot E$ vs. time for \textbf{(a)} Phase-sum-based dynamics. $\dot E$ remains non-positive over the displayed interval. \textbf{(b)} Phase-difference-based dynamics. $\dot E$ repeatedly takes positive values indicating violation of gradient-descent. A randomly generated, 50-node bipartite graph with symmetric antiferromagnetic weights is considered in the analysis.}
    \label{fig:oim_dim_energy_rate}
\end{figure}

\noindent \textit{Gradient structure in the presence of even components}---
We next examine how the phase-sum-based and phase-difference-based dynamics affect energy evolution in an oscillator network. We simulate these two distinct dynamics utilizing an interaction function that contains first-harmonic sine and cosine components, with the cosine term introducing even interactions.
\begin{equation}
\begin{aligned}
    H(\theta_i+\theta_j)
    &=K_{o}\sin(\theta_i+\theta_j)+K_{e}\cos(\theta_i+\theta_j),\\
    H(\theta_i-\theta_j)
    &=K_{o}\sin(\theta_i-\theta_j)+K_{e}\cos(\theta_i-\theta_j),
\end{aligned}
    \label{eq:H_even_interaction}
\end{equation}
here $K_{\mathrm{o}}$ and $K_{\mathrm{e}}$ are the coefficients of the odd and even components, respectively.

Figure~\ref{fig:oim_dim_energy_rate} shows the evolution of $\dot E$ as a function of time for the two interaction laws shown above. We consider a randomly generated bipartite graph with 50 nodes, symmetric weights, and antiferromagnetic coupling. The corresponding candidate energies and their evolution with time are derived in Supplementary Note~5. It can be observed that the energy rate, $\dot E$, of the phase-sum-based dynamics remains non-positive throughout the system evolution. In contrast, the phase-difference-based dynamics exhibit several time intervals in which the $\dot E>0$, further corroborating the violation of pure gradient-flow dynamics.

This analysis establishes that phase-sum interactions preserve gradient flow without the odd-symmetry constraint required by phase-difference coupling. However, this theoretical guarantee applies only when the interaction depends exclusively on the phase-sum. In physical oscillator networks, nonlinear oscillators and their interactions can generate interaction terms that depend on  combinations of oscillator phases beyond phase-sum, thus breaking the gradient-flow structure. Therefore, realizing gradient flow in oscillator hardware requires not only generating the desired phase-sum interaction but also suppressing unwanted harmonic-mixing contributions. In the following section, we examine how these requirements can be achieved.

\subsection{Physical Realization of Phase-Sum Interactions}
\label{sec:harmonic_mixing}

\begin{figure}[!h]
    \centering

    \includegraphics[width=0.99\linewidth]{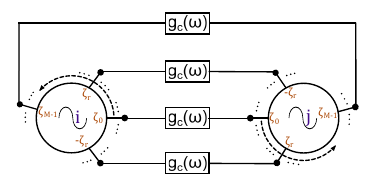}
    \caption{\justifying \small
    \textbf{Phase-resolved coupling architecture for FM phase-sum interactions.} 
    Schematic of two oscillators coupled through M phase-resolved paths. Each path connects taps with complementary phase offsets 
    \(+\zeta\) and \(-\zeta\) 
    through pump-modulated coupling element \(g_{ij}(t)\). The contributions from all paths are combined to suppress cross-harmonic interactions \((n\neq m)\) while retaining the desired phase-sum terms \((n=m)\).
    }
\label{fig:DIM_scheme}
\end{figure}

\noindent We now present our proposed approach to implement phase-sum coupling. Let
\begin{equation}
    \phi_i(t)=\omega t+\theta_i,
    \qquad
    \phi_j(t)=\omega t+\theta_j,
\end{equation}
where $\theta_i$ and $\theta_j$ denote the slowly varying phases. We consider a phase-resolved coupling path from oscillator $j$ to oscillator $i$, in which the receiving tap of oscillator $i$ and the transmitting tap of oscillator $j$ are selected such that the corresponding iPRC and transmitted waveform have complementary phase offsets $+\zeta$ and $-\zeta$, respectively (Fig.~\ref{fig:DIM_scheme}).  Furthermore, the coupling path incorporates an active element whose time-varying conductance exhibits a dominant second-harmonic (\(2\omega\)) component. The implementation details of the complementary offsets using time-shifted oscillator taps as well as the coupling element is discussed in Section~\ref{Phase-Domain Sampling}. 

For the coupling architecture proposed above, the corresponding phase perturbation is
\begin{equation}
    \Gamma_{ij}(t;\zeta)
    =
    g_{ij}(t)\,
    Z(\phi_i+\zeta)\,
    q(\phi_j-\zeta),
    \label{eq:generic_phase_resolved_path}
\end{equation}
where \(Z\) is the iPRC of the receiving oscillator $i$, \(q\) is the waveform transmitted from oscillator \(j\), and \(g_{ij}(t)\) denotes the coupling modulation.

For general nonlinear oscillators and coupling elements, all three
quantities may contain multiple harmonics. We therefore write
\begin{equation}
    Z(\phi)=\sum_{n} Z_n e^{in\phi},
    \qquad
    q(\phi)=\sum_{m} Q_m e^{im\phi},
\end{equation}
and express the coupling modulation, referenced to the second-harmonic
pump, as
\begin{equation}
    g_{ij}(t)
    =
    \sum_k G_k
    e^{ik(2\omega t+\psi_{ij})}.
    \label{eq:coupling_modulation_fourier}
\end{equation}
Here, \(\psi_{ij}\) is the phase of the  applied pump signal to coupling elements. We assume all the coupling elements among oscillator pairs share the same phase pump with identical phase reference $\left(\psi_{ij}=\psi\right)$ without loss of generality. Substitution into Eq.~\eqref{eq:generic_phase_resolved_path} gives
\begin{align}
    \Gamma_{ij}(t;\zeta)
    =
    \sum_{n,m,k}
    G_k Z_n Q_m
    \exp\Big\{i\big[
        &(n+m+2k)\omega t \nonumber \\+n\theta_i+m\theta_j
        &+(n-m)\zeta
        +k\psi
    \big]\Big\}. \label{eq:general_harmonic_mixing}
\end{align}
Since the phase dynamics evolve on a timescale much slower than the fundamental oscillation period, we average the perturbation over one oscillator period to obtain the long term evolution of the oscillator phases. By the orthogonality of Fourier harmonics, all terms with nonzero temporal frequencies integrate to zero, leaving only the temporal resonant contributions satisfying
\begin{equation}
    n+m+2k=0.
    \label{eq:temporal_resonance}
\end{equation}
Here, $n,m,k \in \mathbb{Z}$. While Eq.~\eqref{eq:temporal_resonance}  enforces temporal
frequency matching, it does not guarantee phase-sum interaction. A
harmonic contribution depends purely on $n(\theta_i+\theta_j)$ only when
the phase indices of the two oscillators are equal,
\begin{equation}
    m=n.
    \label{eq:harmonic_matching_condition}
\end{equation}
To ensure phase-sum (only) dynamics, we combine Eqs.~\eqref{eq:temporal_resonance} and
\eqref{eq:harmonic_matching_condition} to give
\begin{equation}
    n=m=-k,
\end{equation}
and hence the desired contribution
\begin{equation}
    \overline{\Gamma}_{ij}^{(+)}
    =
    \sum_n
    G_{-n}Z_nQ_n
    e^{in(\theta_i+\theta_j-\psi)}.
    \label{eq:matched_phase_sum_channel}
\end{equation}
Thus, every matched harmonic contributes through the phase-sum alone, and the modulation spectrum $G_{-n}$ determines which of these harmonics participate in the effective interaction.

The temporal resonance condition (Eq. \ref{eq:temporal_resonance}) on its own also permits terms with $m\neq n$. These cross-harmonic channels generate phase dependences that cannot, in general, be expressed as a function of $n(\theta_i+\theta_j)$. For example, for the fundamental component of the $2\omega$ modulation ($k=1$), the combinations
\begin{equation}
    (n,m,k)\in\left\{
    (1,-3,1),\,(-3,1,1)
    \right\}
\end{equation}
both satisfy Eq.~\eqref{eq:temporal_resonance} and produce contributions
proportional to
\begin{equation}
    e^{i(\theta_i-3\theta_j+4\zeta+\psi)},
    \qquad
    e^{i(-3\theta_i+\theta_j-4\zeta+\psi)},
    \label{eq:mixed_harmonic_examples}
\end{equation}
respectively. Such terms do not align with the phase-sum form required to guarantee gradient-flow and can, in general, introduce non-conservative components into the reduced phase dynamics. The above analysis indicates that any physical realization of phase-sum dynamics must retain 
the matched-harmonic channels $n=m$ while eliminating the cross-harmonic channels $n\neq m$.

Importantly, Eq.~\eqref{eq:general_harmonic_mixing} also reveals that these two categories of harmonic channels are distinguished by their dependence on the phase offset
$\zeta$. Each $(n,m)$ interaction acquires the factor
\( e^{i(n-m)\zeta}.\)  The desired matched-harmonic channel $n=m$ is therefore the zero-frequency component with respect to the phase coordinate $\zeta$ whereas every cross-harmonic channel $n\neq m$ oscillates along this coordinate.

This observation provides a basis for selectively retaining the desired phase-sum interactions through averaging over the phase offset $\zeta$. Since matched-harmonic terms with $n=m$ are independent of $\zeta$, they remain unchanged under averaging, whereas cross-harmonic terms with $n\neq m$ can be eliminated. In the following section, we first establish this result using continuous phase-domain averaging and then examine how it can be realized using a finite number of phase samples.

\subsection{Nyquist-like Criterion to Achieve Pure Phase Sum Interactions}
\label{sec:phase_sampling}

\noindent The key observation from the preceding section is that a term involving the $n^{th}$ harmonic of the iPRC and the $m^{th}$ harmonic of the transmitted waveform carries a phase-offset dependence of $e^{i(n-m)\zeta}$. This suggests combining coupling paths with different values of $\zeta$ so that the cross-harmonic terms cancel while the matched-harmonic terms add constructively.\\

\noindent \textit{Coupling with a continuum of phase offsets}---We first consider an idealized construction in which coupling paths with a continuum of phase offsets, $\zeta\in[0,2\pi)$ are available. Their contributions
are combined with equal weight,
\begin{equation}
    \overline{\Gamma}^{\,\zeta}_{ij}(t)
    =
    \frac{1}{2\pi}
    \int_{0}^{2\pi}
    \Gamma_{ij}(t;\zeta)\,d\zeta.
    \label{eq:continuous_phase_average}
\end{equation}
Substituting the harmonic expansion derived in
Eq.~\eqref{eq:general_harmonic_mixing}, the dependence on $\zeta$ is
contained entirely in
\[
    \frac{1}{2\pi}
    \int_{0}^{2\pi}
    e^{i(n-m)\zeta}\,d\zeta
    =
    \begin{cases}
        1, & n=m,\\
        0, & n\neq m.
    \end{cases}
  \]
Hence, combining coupling paths continuously over the phase offset eliminates every term involving unequal harmonic orders $n\neq m$ and retains only the matched terms $n=m$.

\noindent Together with the temporal resonance condition
\begin{equation}
    n+m+2k=0,
\end{equation}
the condition $n=m$ gives $k=-n$. The resulting slow interaction is
therefore
\begin{equation}
    \overline{\Gamma}^{(\infty)}_{ij}
    =
    \sum_n
    G_{-n}Z_nQ_n
    e^{in(\theta_i+\theta_j-\psi)}
    \equiv
    H(\theta_i+\theta_j-\psi).
    \label{eq:continuum_phase_sum}
\end{equation}
Thus, averaging over all  coupling paths with an (idealized) continuum of phase offsets removes all the
cross-harmonic mixing responsible for terms (e.g., 
$\theta_i-3\theta_j$), while retaining all matched harmonics as functions
of the phase-sum. If the coupling modulation harmonic is only confined to the fundamental component, i.e., $|k|=1$, then the interaction in Eq.~\eqref{eq:continuum_phase_sum} represents the exact DIM dynamics.\\

\noindent \textit{Finite phase-resolved coupling paths and harmonic aliasing}---While the continuum formulation above provides an idealized framework for implementing any integrable phase-sum based function ($n(\theta_i+\theta_j)$) for oscillator interaction, practical architectures will be restricted to a finite number ($M$) of coupling pathways,  
each with a given phase  offset
\(
    \zeta_0,\,\zeta_1,\,\ldots,\,\zeta_{M-1}.
\)
Consider \(M\) uniformly spaced phase offsets over the interval
\(\zeta\in[0,2\pi)\),
\begin{equation}
    \zeta_r=\frac{2\pi r}{M},
    \qquad r=0,\ldots,M-1.
\end{equation}
The combined interaction can then expressed as
\begin{equation}
    \overline{\Gamma}^{(M)}_{ij}(t)
    =
    \frac{1}{M}
    \sum_{r=0}^{M-1}
    \Gamma_{ij}(t;\zeta_r).
    \label{eq:finite_path_average}
\end{equation}
The finite set of
offsets $\zeta_r$ constitutes a discrete sampling of the continuous
phase-offset coordinate $\zeta$. For a term involving harmonic orders $(n,m)$, averaging over the $M$ paths gives
the factor
\begin{align}
    A_M(n,m)
    &=
    \frac{1}{M}
    \sum_{r=0}^{M-1}
    e^{i(n-m)\zeta_r}
    \nonumber\\
    &=
    \frac{1}{M}
    \sum_{r=0}^{M-1}
    e^{i2\pi r(n-m)/M}.
    \label{eq:finite_path_factor}
\end{align}
Discrete Fourier orthogonality then gives
\begin{equation}
    A_M(n,m)
    =
    \begin{cases}
        1, & n-m=\ell M,\quad \ell\in\mathbb{Z},\\[3pt]
        0, & \text{otherwise}.
    \end{cases}
    \label{eq:aliasing_condition}
\end{equation}
The desired terms with $n=m$ therefore continue to add constructively. However, unlike the continuous average, a finite number of coupling paths also retain certain cross-harmonic terms with $n\neq m$. Specifically, harmonic orders separated by an integer multiple of $M$ produce the same values at all $M$ phase offsets and therefore cannot be distinguished by the finite set of paths. This phenomenon can be considered as \emph{aliasing} in the phase domain, analogous to aliasing in frequency sampling where the sample count below a certain minimum fails to reconstruct a signal with accurate frequency. Similarly, when the number of coupling paths is below a certain threshold, the resulting phase interaction fails to eliminates cross harmonic mixing. In the following analysis, we adopt a ``Nyquist-like'' sampling approach to derive the minimum number of coupling paths required to eliminate cross-harmonic interactions within a specific harmonic range.\\

\noindent \textit{Nyquist-like criterion for the number of phase-resolved coupling paths}
\label{subsec:nyquist_phase_sampling}
---The aliasing condition derived above directly determines how many phase-resolved coupling paths are required to eliminate all cross-harmonic terms over a specified harmonic range.

Let \(N\) denote the highest harmonic order present in either the phase
response or the transmitted waveform,
\begin{equation}
    N=\max\left\{|n|:Z_n\neq0\ \text{or}\ Q_n\neq0\right\}.
\end{equation}
For any pair of participating harmonic orders \(n\) and \(m\),
\begin{equation}
    |n-m|\leq 2N.
\end{equation}
Since aliasing occurs when
\begin{equation}
    n-m=\ell M,
    \qquad \ell\in\mathbb{Z}\setminus\{0\},
\end{equation}
choosing \(M>2N\) ensures that no nonzero harmonic-order difference can
satisfy the aliasing condition. Thus, the smallest integer number of
phase-resolved FM coupling paths that guarantees alias-free harmonic
selection based solely on the highest harmonic order \(N\) is
\begin{equation}
    M_{\mathrm{cert}}=2N+1.
    \label{eq:general_minimum_paths}
\end{equation}


We also consider oscillators whose waveforms and corresponding iPRCs exhibit half-wave antisymmetry,
\begin{equation}
    q(\phi+\pi)=-q(\phi),
    \qquad
    Z(\phi+\pi)=-Z(\phi),
    \label{eq:half_wave_symmetry_sampling}
\end{equation}
such that only odd harmonics are present. This odd-harmonic structure also enables the sign reversal of the interaction under a $\pi$ phase shift, as required for implementing signed Ising interactions~\cite{2026arXiv260302933H}.

Since $n$ and $m$ are both odd, their difference $(n-m)$ is always even, making the phase-offset dependence $e^{i(n-m)\zeta}$ periodic over $\pi$ rather than $2\pi$.

Consequently, for an even number of uniformly spaced physical taps $M$, coupling paths with phase offsets separated by $\pi$ produce identical contributions. Thus, only $M/2$ independent coupling paths are required, each connecting complementary phase offsets $+\zeta_r$ and $-\zeta_r$ of oscillators $i$ and $j$, respectively, where
\begin{equation}
\zeta_r=\frac{2\pi r}{M},
\qquad r=0,\ldots,\frac{M}{2}-1.
\label{eq:independent_tap_offsets}
\end{equation}
The negative offsets are interpreted modulo $2\pi$. For example, $+90^\circ$ on oscillator $i$ is coupled to $-90^\circ$ ($270^\circ$) on oscillator $j$, as illustrated in Fig.~\ref{fig:DIM_scheme}. Therefore, the number of independent phase-resolved coupling paths is
\begin{equation}
M_{\mathrm{paths}}=\frac{M}{2}.
\end{equation}


Under this sampling convention, cross-harmonic terms survive only when
\begin{equation}
n-m=2\ell M_{\mathrm{paths}},
\qquad \ell\in\mathbb{Z}.
\label{eq:independent_path_aliasing}
\end{equation}
For an oscillator containing the first $P$ odd harmonics,
\begin{equation}
n,m\in{\pm1,\pm3,\ldots,\pm(2P-1)},
\end{equation}
the maximum harmonic-order difference is $|n-m|_{\max}=4P-2$. Avoiding phase-domain aliasing therefore requires
\begin{equation}
2M{\mathrm{paths}}>4P-2,
\end{equation}
yielding
\begin{equation}
M_{\mathrm{paths},\min}=2P.
\label{eq:minimum_phase_paths}
\end{equation}

For oscillator Ising machines, half-wave antisymmetry additionally enables the realization of both FM and AFM interactions. Since only odd harmonics contribute, shifting the transmitting tap by $\pi$ reverses the sign of the phase-sum interaction. An even number of uniformly spaced physical taps ensures the availability of diametrically opposite tap pairs, enabling both coupling polarities. Consequently, the minimum number of physical taps required to support both FM and AFM interactions is the smallest even number that satisfies the criterion for eliminating phase domain aliasing.

Equation~\eqref{eq:minimum_phase_paths} yields a ``Nyquist-like'' criterion for phase-resolved oscillator coupling guaranteeing rejection of cross-harmonic channels within a particular harmonic bandwidth. In the subsequent sections, we will simulate and show this effect as well as confirm that within a given bandwidth phase-sum dynamics, even when realized with finite number of oscillator taps, can provide better alignment with gradient flow dynamics compared to phase-difference based conventional OIMs.

\begin{figure*}
    \centering
    \includegraphics[width=0.99\linewidth]{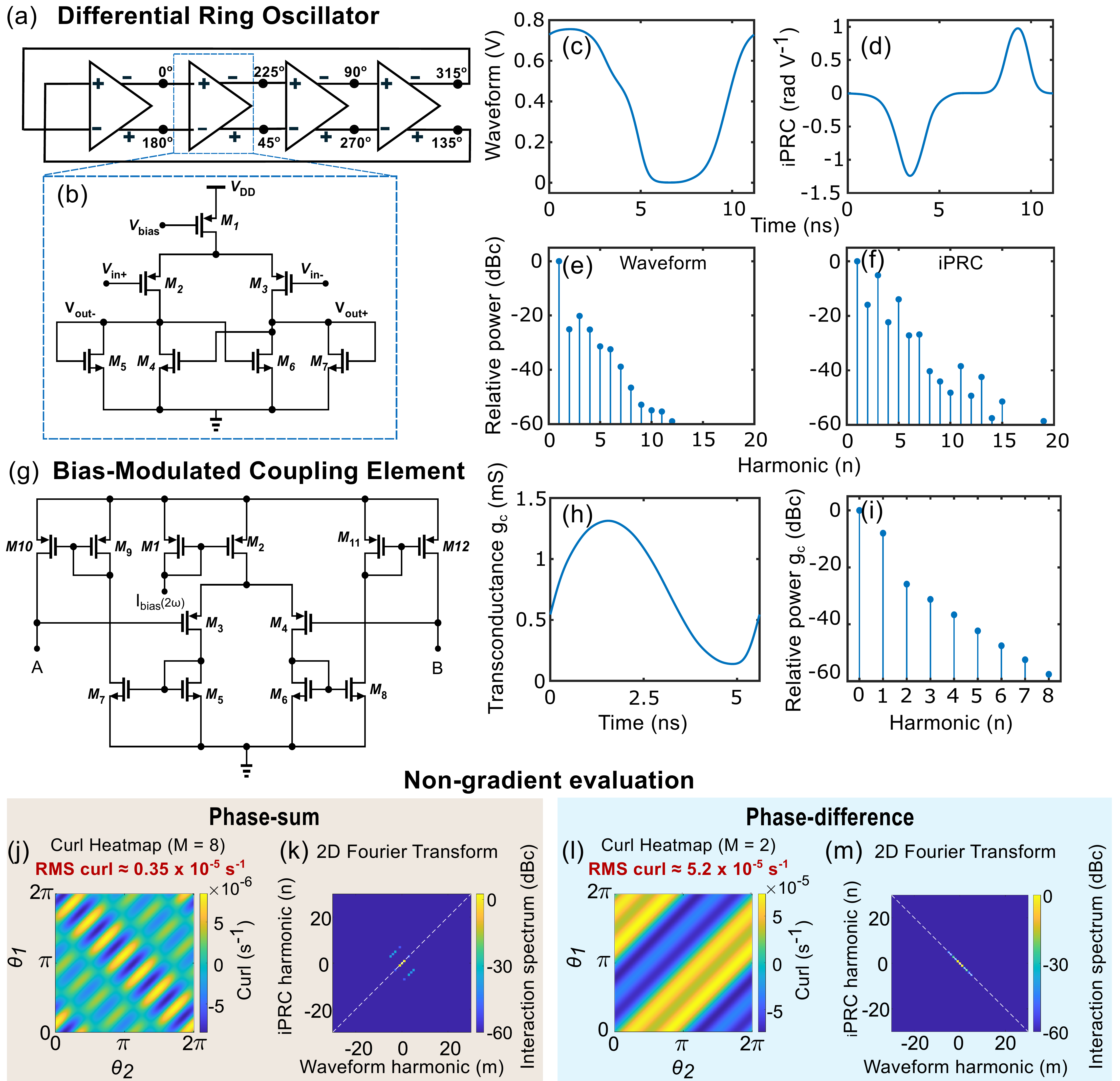}
    \caption{\justifying{\small\textbf{Comparison of non-gradient curl in phase-sum and phase-difference oscillator coupling.}
\textbf{(a)} Four-stage differential ring oscillator architecture providing eight taps with phase offsets in increments of $45^\circ$ relative to the $0^\circ$ reference.
\textbf{(b)} Schematic of the differential delay cell used in each stage.
\textbf{(c,d)} Voltage waveform and corresponding iPRC at the reference tap over one oscillation period.
\textbf{(e,f)} Harmonic power spectra of the reference waveform and iPRC, respectively.
\textbf{(g)} Schematic of the bias-modulated coupling element, with its transconductance controlled by $I_{\mathrm{bias}}$.
\textbf{(h,i)} Time-varying coupling transconductance under second-harmonic modulation and its corresponding harmonic spectrum, including the DC component. The spectra in (e), (f), and (i) are displayed with a $-60\,\mathrm{dBc}$ cutoff relative to their respective reference spectral components.
\textbf{(j)} Numerically evaluated curl of the eight-path phase-sum coupling dynamics over the two-dimensional phase space $(\theta_1,\theta_2)$.
\textbf{(k)} Corresponding two-dimensional harmonic spectrum, showing the desired phase-sum contributions $n(\theta_1+\theta_2)$ along the diagonal ($n=m$) and residual cross-harmonic contributions $n\theta_1+m\theta_2$ ($n\neq m$) arising from phase-domain aliasing along the off-diagonal.
\textbf{(l,m)} Corresponding curl distribution and harmonic spectrum for conventional phase-difference coupling, where the harmonic contributions are concentrated along the anti-diagonal ($n=-m$), representing interactions of the form $n(\theta_1-\theta_2)$.
The RMS curl is $0.35\times10^{-5}\,\mathrm{s}^{-1}$ for phase-sum coupling and $5.2\times10^{-5}\,\mathrm{s}^{-1}$ for conventional phase-difference coupling, demonstrating an approximately fifteenfold reduction in non-gradient contributions with the proposed phase-resolved coupling architecture.}}
    \label{fig:DIM_OIM_curl}
\end{figure*}

\subsection{Circuit Implementation}
\label{Phase-Domain Sampling}
\noindent We next investigate the physical realization of phase-sum interactions and evaluate the extent to which the predicted phase-sum dynamics and gradient-flow properties are preserved in practical circuits. Guided by the phase-resolved coupling framework developed above, we consider a four-stage differential ring oscillator providing eight phase-shifted taps, together with a bias-modulated coupling element. The oscillator architecture and coupling circuit are described below, followed by their numerical evaluation using transistor-level simulations in Cadence Virtuoso with TSMC 65-nm CMOS technology. \\

\noindent\textit{Differential Ring Oscillator Architecture}---The block diagram of the four-stage differential ring oscillator is shown in Fig.~\ref{fig:DIM_OIM_curl}(a). Each stage consists of a differential delay cell as shown in Fig.~\ref{fig:DIM_OIM_curl}(b), where \(V_{\mathrm{bias}}\) controls the stage current, and thus tunes the free-running oscillation frequency. Fig.~\ref{fig:DIM_OIM_curl}(c) shows the voltage waveform at the reference tap over one oscillation period. The four-stage differential ring oscillator nominally provides eight equally spaced phase taps, spanning a full oscillation cycle in increments of $45^\circ$ ($\pi/4$). Fig.~\ref{fig:DIM_OIM_curl}(d) shows the corresponding iPRC at the reference tap, while the iPRCs of the remaining taps can be obtained through the respective phase shifts. The harmonic spectra of the reference waveform and its iPRC are presented in Figs.~\ref{fig:DIM_OIM_curl}(e) and (f), respectively. In both spectra, the harmonic power is normalized to that of the fundamental component, and only harmonics within $60$ dB of the fundamental ($\geq -60$ dBc) are retained for subsequent analysis. \\
\noindent\textit{Bias-Modulated Coupling Element---} Each coupling path is implemented using the variable transconductance element shown in Fig.~\ref{fig:DIM_OIM_curl} (g). Its two oscillator-facing terminals, \(A\) and \(B\), are connected to the selected taps of oscillators \(i\) and \(j\) as in Fig.~\ref{fig:DIM_scheme}, while the bias terminal is driven by
\begin{equation}
    I_{\mathrm{bias}}(t)
    =
    I_0+I_1\cos\!\left(2\omega t+\psi\right),
    \qquad I_0\geq |I_1|.
    \label{eq:ibias_current}
\end{equation}
Here, \(I_0\) sets the DC operating point, \(I_1\) controls the modulation amplitude of the pump, and \(\psi\) is the modulation phase of the pump signal. The circuit senses the voltage difference between ports \(A\) and \(B\) through a differential pair and returns mirrored currents to the same ports. The port-current relations and the dependence of the effective coupling transconductance, $g_c(t)$, on $I_{\mathrm{bias}}$, including the corresponding measurements, are derived in Supplementary Note~6.
Fig.~\ref{fig:DIM_OIM_curl}(h) shows the corresponding pump modulated transconductance, \(g_c(t)\), while Fig.~\ref{fig:DIM_OIM_curl}(i) shows the harmonic content of \(g_c(t)\) under periodic modulation via \(I_{\mathrm{bias}}(t)\). It can be observed that the response is dominated by the \(2\omega\) component, which accounts for approximately \(97.7\%\) of the mean-square AC content of the first eight harmonics retained beyond the $-60$ dBc cutoff. 

To realize FM phase-sum coupling between oscillators $i$ and $j$, we select eight coupling paths with complementary phase offsets, following the phase-resolved coupling scheme developed in Section D. Specifically, the tap connections are:
\[
\begin{aligned}
&0^\circ \leftrightarrow 0^\circ, \qquad
\,\,\,\,\,\,\,\,\,\,\,\,45^\circ \leftrightarrow 315^\circ,\\
&90^\circ \leftrightarrow 270^\circ, \qquad
\,\,135^\circ \leftrightarrow 225^\circ,\\
&180^\circ \leftrightarrow 180^\circ, \qquad
225^\circ \leftrightarrow 135^\circ,\\
&270^\circ \leftrightarrow 90^\circ, \qquad \,\,\,
315^\circ \leftrightarrow 45^\circ.
\end{aligned}
\]
Each path incorporates the bias-modulated coupling element described above, with the contributions from all eight paths combined to realize the desired phase-sum interaction. As established by the phase-domain sampling analysis, this configuration retains matched-harmonic contributions of the form $n(\theta_i+\theta_j)$ while suppressing cross-harmonic terms $n\theta_i+m\theta_j$ for which $n\neq m$. However, because the number of coupling paths is finite, cross-harmonic components satisfying the aliasing condition $n-m=8\ell$, where $\ell$ is a nonzero integer, can survive. These residual terms introduce phase dependences beyond the desired phase-sum interaction and can generate non-conservative contributions to the reduced phase dynamics.

To quantify the resulting deviation from gradient flow, we numerically evaluate the curl of the phase dynamics over the two-dimensional phase space $(\theta_1,\theta_2)$. Fig.~\ref{fig:DIM_OIM_curl}(j) shows the resulting curl distribution for the eight-path phase-sum coupling configuration. The corresponding two-dimensional harmonic spectrum is presented in Fig.~\ref{fig:DIM_OIM_curl}(k), where the diagonal components ($n=m$) represent the desired phase-sum interactions, while the off-diagonal components ($n\neq m$) identify the residual cross-harmonic contributions associated with phase-domain aliasing.

For comparison, we also evaluate conventional phase-difference coupling using the same oscillator architecture. Here, two FM coupling paths connect the corresponding $0^\circ\leftrightarrow0^\circ$ and $180^\circ\leftrightarrow180^\circ$ taps, with the coupling transconductance held constant rather than modulated at $2\omega$. Consequently, only the zeroth-order modulation harmonic contributes, and temporal averaging retains interactions of the form $n(\theta_1-\theta_2)$. Unlike the phase-sum configuration, no phase-domain aliasing arises from the coupling-path discretization. Nevertheless, as established in Section A, the resulting phase-difference interaction is not guaranteed to preserve gradient flow because of harmonic misalignment between the oscillator waveform and its iPRC. Fig.~\ref{fig:DIM_OIM_curl}(l) shows the corresponding curl distribution, while Fig.~\ref{fig:DIM_OIM_curl}(m) confirms that the interaction spectrum is concentrated along the anti-diagonal ($n=-m$), characteristic of phase-difference coupling.

To quantitatively compare the two coupling configurations, we calculate the root-mean-square (RMS) curl over the phase space (see Supplementary Note 7). The phase-sum configuration yields an RMS curl of $0.35\times10^{-5}\,\mathrm{s}^{-1}$, compared with $5.2\times10^{-5}\,\mathrm{s}^{-1}$ for conventional phase-difference coupling, corresponding to an approximately fifteenfold reduction. Although the finite number of phase-resolved coupling paths does not completely eliminate the non-gradient contributions arising from phase-domain aliasing, the substantially reduced curl demonstrates that the proposed architecture more closely approximates gradient-flow dynamics than conventional phase-difference coupling for the oscillator and coupling configurations considered.

\begin{figure*}
    \centering
    \includegraphics[width=\textwidth]{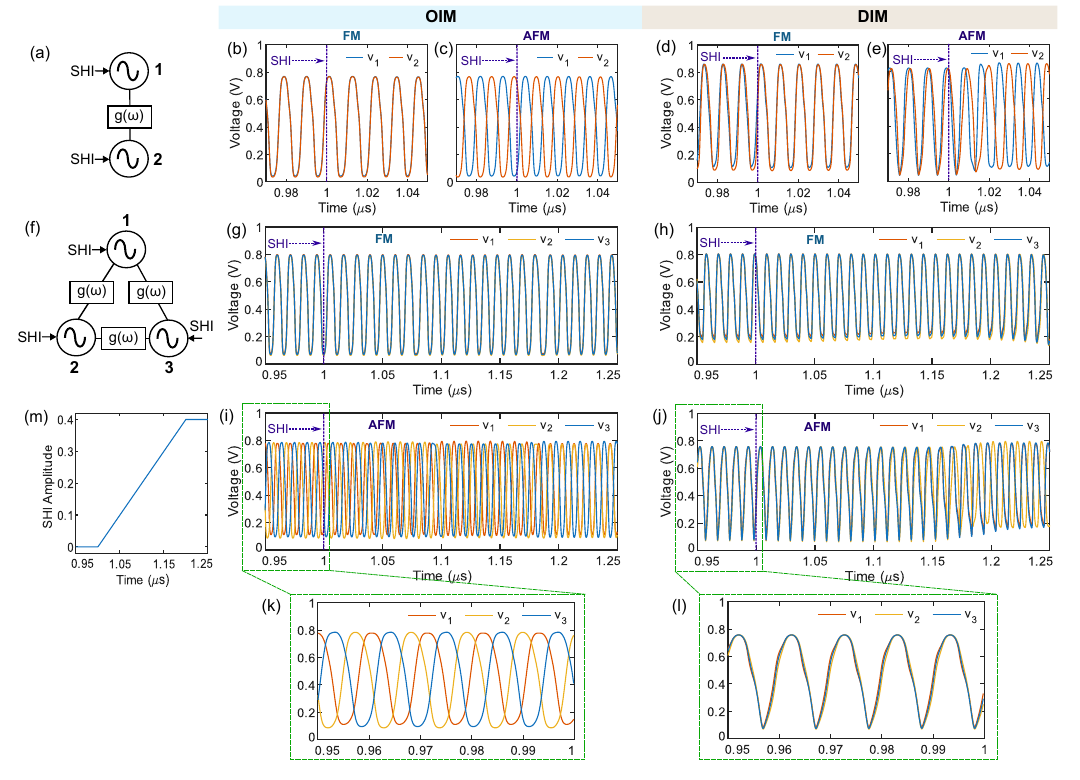}
    \caption{\justifying \small\textbf{Comparison of phase dynamics in conventional OIMs and phase-sum-based DIMs.}
\textbf{(a)} Two-oscillator network with symmetric coupling.
\textbf{(b,c)} OIM responses under FM and AFM coupling, exhibiting in-phase and opposite-phase synchronization, respectively.
\textbf{(d,e)} Corresponding DIM responses. Under FM coupling, the oscillators synchronize in phase, whereas under AFM coupling, they initially converge to a common-phase fixed point at $\pi/2$ (relative to the pump-phase reference) before separating into opposite binary phases upon application of SHI.
\textbf{(f)} Fully connected three-oscillator network.
\textbf{(g,h)} OIM and DIM responses, respectively, under all-to-all FM coupling, showing in-phase synchronization in both systems.
\textbf{(i,j)} Corresponding responses under all-to-all AFM coupling. Before SHI, the OIM exhibits a frustrated configuration with approximately $2\pi/3$ phase separation, whereas the DIM synchronizes all three oscillators to a common phase.
\textbf{(k,l)} Magnified views of (i) and (j), respectively, highlighting the distinct synchronization states before SHI.
\textbf{(m)} Temporal profile of the applied SHI amplitude. The vertical dashed lines in the waveform panels indicate the onset of the SHI ramp at $t=1\,\mu\mathrm{s}$.
    }
    \label{fig:small_network_diagnostics}
\end{figure*}

\subsection{Implementing DIMs}
\label{sectionV}
\noindent

The Nyquist-like criterion derived in Section~\ref{subsec:nyquist_phase_sampling} establishes that two phase-resolved coupling paths are sufficient to realize the fundamental phase-sum interaction for oscillators exhibiting half-wave antisymmetry. Nevertheless, we employ four coupling paths to ensure balanced loading of the oscillator quadratures. For the differential ring oscillator shown in Fig.~\ref{fig:DIM_OIM_curl}(a), the tap connections between oscillators $i$ and $j$ are selected as
\[
0^\circ \leftrightarrow 0^\circ,\quad
90^\circ \leftrightarrow 270^\circ,\quad
180^\circ \leftrightarrow 180^\circ,\quad
270^\circ \leftrightarrow 90^\circ.
\]
As established by Eq.~\eqref{eq:general_harmonic_mixing}, the phase-resolved coupling scheme retains matched-harmonic interactions of the form $H(\theta_i+\theta_j)$. For the DIM implementation, we consider sine harmonics,
\begin{equation}
H(\theta_i+\theta_j)
=
\sum_{\substack{n\geq 1\\ n\,\mathrm{odd}}}
h_n\sin\left[n(\theta_i+\theta_j)\right],
\end{equation}
which preserve gradient-flow dynamics and the Ising energy mapping on the binary phase manifold. Retaining the dominant fundamental harmonic, the interaction reduces to $H(\theta_i+\theta_j)\propto\sin(\theta_i+\theta_j)$. The corresponding idealized DIM dynamics for a network of oscillators with symmetric coupling weights $W_{ij}$ are given by
\begin{equation}
    \dot{\theta}_i
    =
    -K\sum_{j\neq i}W_{ij}\sin(\theta_i+\theta_j).
    \label{eq:network_phase_sum_only}
\end{equation}
\noindent Furthermore, when SHI is additionally applied to the readout port of the oscillator, Eq.~\eqref{eq:network_phase_sum_only} evolves to,
\begin{equation}
    \dot{\theta}_i
    =
    -K\sum_{j\neq i}W_{ij}\sin(\theta_i+\theta_j)
    -
    K_s\sin(2\theta_i).
    \label{eq:implemented_dim_full}
\end{equation}
which corresponds to dynamics of the DIM introduced in Eq.~\eqref{eq:intro_dim}. The dynamics in Eq.~\eqref{eq:implemented_dim_full} are a gradient flow of the scalar potential
\begin{equation}
    E(\boldsymbol{\theta})
    =
    -K\sum_{i<j}W_{ij}\cos(\theta_i+\theta_j)
    -
    \frac{K_s}{2}\sum_i\cos(2\theta_i).
    \label{eq:implemented_dim_potential}
\end{equation}
At the binary phase states \(\theta_i \in \{0,\pi\}\), the corresponding Ising spins are defined as $\sigma_i=\cos\theta_i\in{-1,+1}$. Consequently,
\begin{equation}
    \cos(\theta_i+\theta_j)
    =
    \sigma_i\sigma_j, \quad \text{and} \quad \cos(2\theta_i)=1.
\end{equation}

\noindent Therefore, on the binary phase manifold,
\begin{equation}
    E(\boldsymbol{\sigma})
    =
    -K\sum_{i<j}W_{ij}\sigma_i\sigma_j
    -
    \frac{NK_s}{2}.
    \label{eq:implemented_dim_binary_energy}
\end{equation}
Thus, the dynamics map to the target Ising Hamiltonian. \\

\noindent \textit{Circuit simulation results}---
We next evaluate the DIM dynamics through circuit-level simulations of coupled oscillator networks. We begin with two- and three-oscillator configurations under FM and AFM coupling, comparing the responses of DIMs and conventional OIMs to highlight the distinct dynamical behavior arising from phase-sum and phase-difference interactions. We then extend the analysis to an interconnected network of 15 oscillators to investigate the collective phase dynamics and the emergence of binary spin states under second-harmonic injection.

\begin{figure*}[t]
    \centering
    \includegraphics[width=\textwidth]{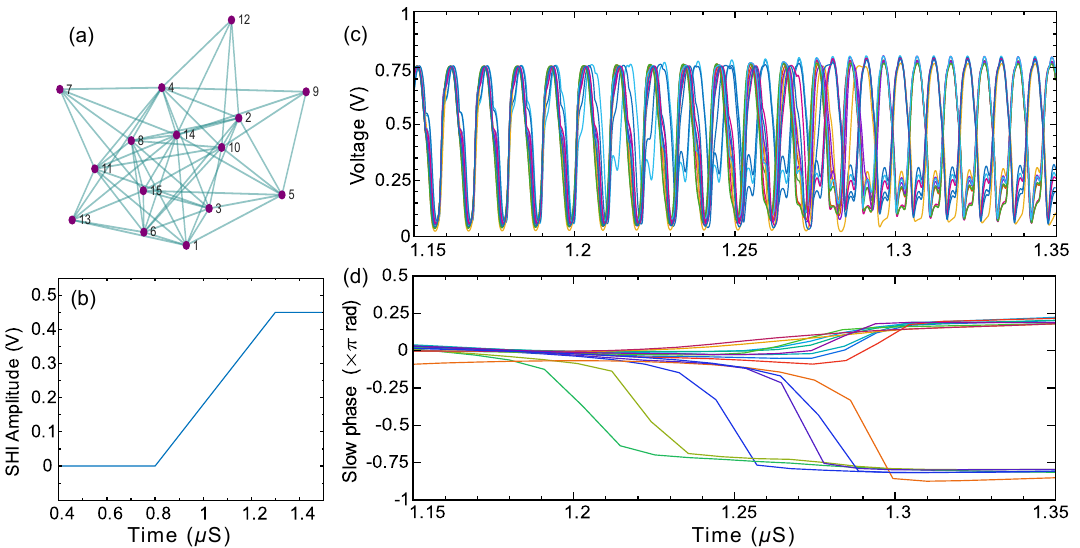}
    \caption{\justifying\small
\textbf{Circuit-level demonstration of a 15-node dynamical Ising machine.}
\textbf{(a)} Randomly generated graph comprising 15 nodes and 59 edges, with each node implemented using a differential ring oscillator and each edge realizing AFM phase-sum coupling. Four taps are considered.
\textbf{(b)} Temporal profile of the applied second-harmonic injection (SHI) amplitude.
\textbf{(c)} Simulated output voltage waveforms of the 15 coupled oscillators.
\textbf{(d)} Corresponding evolution of the extracted slow oscillator phases. As the SHI amplitude increases, the oscillator phases separate into two groups differing by approximately $\pi$, corresponding to the binary spin states $\sigma_i\in\{-1,+1\}$.}
    \label{fig:graph_simulation}
\end{figure*}

\noindent \subparagraph{Phase Sum Coupling in a Two-Oscillator Network---} To validate the phase-sum dynamics, we first implement a two-oscillator network as shown in  Fig.~\ref{fig:small_network_diagnostics}(a) with FM and AFM interactions. The coupling-modulation pump is applied continuously, whereas the local SHI is ramped from \(t=1~\mu\mathrm{s}\) to confine the oscillator phases to the binary manifold. Its temporal profile is shown in Fig.~\ref{fig:small_network_diagnostics}(m). We also implement the corresponding phase-difference dynamics for reference.

Figs.~\ref{fig:small_network_diagnostics}(b) and \ref{fig:small_network_diagnostics}(c) show the conventional OIM results with phase difference interactions. FM coupling synchronizes the two oscillators in phase, whereas AFM coupling produces a phase difference of approximately \(\pi\) between the oscillators. The corresponding phase-sum-based DIM results are shown in Figs.~\ref{fig:small_network_diagnostics}(d) and \ref{fig:small_network_diagnostics}(e). For FM coupling, the two oscillators again settle to the same phase. For AFM coupling, oscillators initially approach a stable fixed point 
$\theta^\star \in \left\{\frac{\pi}{2}\right\}$
relative to the modulation reference. However, once the SHI strength exceeds a threshold~\cite{cdc9-y234}, the oscillators separate into opposite binary phases, with a relative phase difference of \(\pi\). In contrast, with phase difference dynamics, the oscillators exhibit a phase difference of $\pi$ even without the application of the SHI. The observed behaviors for the OIM and DIM can be explained by analyzing the fixed points of the system and their stability as detailed in Supplementary Note~8. Despite the different transient responses, both implementations yield the same binary phase relationships, reflecting their common Ising representation on the binary manifold.

\subparagraph{Three-Oscillator Network---} Next, we consider a 3 oscillator network with all-to-all coupling as illustrated in Fig.~\ref{fig:small_network_diagnostics}(f), where the phase-sum and phase-difference dynamics produce distinct phase states prior to the application of SHI and consequent convergence to the binary manifold.

With FM coupling, the oscillators synchronize in phase for both DIM and OIM cases, as shown in Figs.~\ref{fig:small_network_diagnostics}(g) and \ref{fig:small_network_diagnostics}(h). This is behavior is expected since $\theta^\star =\{0,0,0\}$ is the most energetically-favorable stable fixed point of the system.

In contrast, for AFM coupling the oscillators exhibit a phase separation of \(2\pi/3\) in the case of the OIM (Figs.~\ref{fig:small_network_diagnostics}(i) and (k)) whereas in the case of the DIM, the oscillators lock-in phase with each other (Figs.~\ref{fig:small_network_diagnostics}(j) and (l)). These distinct characteristics can be explained by analyzing the fixed points. For the OIM, the equally spaced configuration with \(2\pi/3\) phase separation is the lowest energy stable fixed point.  In DIM, each AFM edge instead favors \(\theta_i+\theta_j=\pi\) modulo \(2\pi\), relative to the modulation reference. This condition holds on all three edges at the common-phase fixed point
\[
    \theta_1=\theta_2=\theta_3=\frac{\pi}{2}
    \quad (\mathrm{mod}\ \pi),
\]
relative to the modulation reference. Accordingly, the coincident waveforms in Fig.~\ref{fig:small_network_diagnostics}(l) show mutual in-phase synchronization before SHI is applied.

As SHI is ramped, the common-phase fixed point $\theta^\star \in \left\{\frac{\pi}{2}\right\}$ of the DIM becomes unstable, and the oscillator phases separate toward \(\theta_i\in\{0,\pi\}\). On the binary manifold, the network is frustrated for both interaction laws, and at least one AFM edge remains \emph{unsatisfied}. 

The distinct behaviors of the OIM and DIM prior to application of the SHI provide direct evidence that the two systems realize fundamentally different oscillator coupling laws.

\textbf{15-Node Oscillator Network}---
We then evaluate a DIM using the above simulation framework on the randomly generated 15-node graph shown in Fig.~\ref{fig:graph_simulation}(a). The SHI amplitude is ramped as shown in Fig.~\ref{fig:graph_simulation}(b). The oscillator voltages and extracted slow phases are shown in Figs.~\ref{fig:graph_simulation}(c) and \ref{fig:graph_simulation}(d), respectively. At low SHI amplitude, the oscillators share a nearly common phase, as in the three-oscillator network. As the SHI amplitude increases, the phases separate into two groups approximately \(\pi\) apart, representing the two spin states and defining a partition of the graph. These results demonstrate network-level operation of the proposed architecture, with each oscillator interacting with multiple neighbors according to the phase-sum interaction law.\\

\begin{figure}
    \centering
    \includegraphics[width=0.99\linewidth]{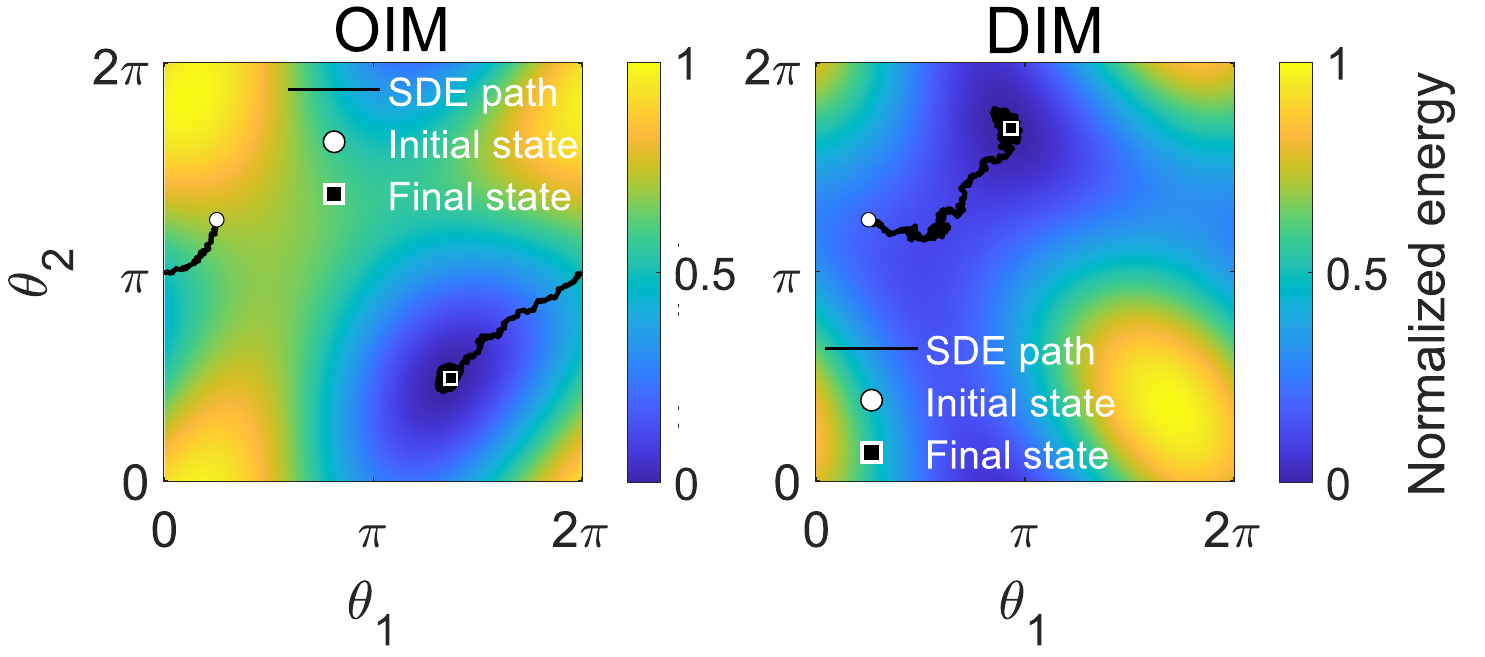}
    \caption{\justifying\small
\textbf{Comparison of the continuous energy landscapes of conventional OIMs and DIMs.}
Conditional energy landscapes for the 15-node graph shown in Fig.~\ref{fig:graph_simulation}(a), obtained by varying the phases $(\theta_1,\theta_2)$ while holding the remaining 13 oscillator phases fixed.
\textbf{(a)} Energy landscape corresponding to conventional phase-difference coupling (OIM).
\textbf{(b)} Corresponding landscape for phase-sum coupling (DIM).
In both cases, $\theta_1$ and $\theta_2$ are swept over $[0,2\pi]$, with color representing the system energy.}
    \label{fig:conditional_energy_landscapes}
\end{figure}

\noindent\textit{Computational implications of different oscillator interaction law}
\label{subsec:graph_dependent_performance}---Although the OIM and DIM encode the same Ising objective at the binary phase configurations, their continuous energy landscapes away from these states are fundamentally different. This is illustrated in Figs.~\ref{fig:conditional_energy_landscapes} (a) and (b), which compares the conditional energy landscapes in the $(\theta_1,\theta_2)$ plane for the OIM and DIM for the 15-node graph considered in Fig.~\ref{fig:graph_simulation} (a),  with the phases of the remaining 13 oscillators held fixed. Thus, the same discrete Ising configurations are embedded within distinct continuous landscapes.

These different embeddings can produce different dynamical properties, including the local stability of the Ising fixed points, as analyzed in our prior work~\cite{cdc9-y234}. Although the DIM dynamics exhibit gradient-flow behavior, the computational benefit of either dynamical model is not universal, but depends on the input graph. 
To illustrate this dependence, we consider four unweighted AFM graph instances, with the Ising objective
\begin{equation}
    E_{\mathrm{Ising}}
    =-\sum_{i<j}W_{ij}\sigma_i\sigma_j,
    \qquad \sigma_i\in\{-1,+1\},
\end{equation}
where \(W_{ij}=-1\) for an edge and zero otherwise. For each graph, 100 paired trials are performed using identical initial phases, noise realizations, simulation times, and parameter schedules, with $K$ held fixed while $K_s$ is gradually increased.

As summarized in Table~\ref{tab:ground_state_performance}, the two phase-interaction laws exhibit markedly different probabilities of reaching the ground state, with neither model consistently outperforming the other across all graph instances. These results demonstrate that changing the interaction law provides access to complementary dynamical searches of the same Ising objective. This suggests an architecture in which multiple oscillator cores implementing distinct interaction laws operate on the same problem instance, allowing the solution to be selected from the outcomes of dynamically distinct searches without requiring \emph{a priori} knowledge of which model is better suited to the particular graph.

\begin{table}[t]
    \caption{\justifying\small
    Comparison of minmum Ising energy obtained using OIM and DIM phase-model simulations. For each graph, \(E^\star\) is the ground state energy, \(E_{\mathrm{best}}\) is the lowest Ising energy obtained by each model, and \(N_{\mathrm{GS}}\) is the number of trials reaching the ground state. Each model is evaluated over 100 trials, with the same initial phases used for each paired OIM and DIM trial.}
    \label{tab:ground_state_performance}
    \centering
    \setlength{\tabcolsep}{3.2pt}
    \begin{ruledtabular}
    \begin{tabular}{ccccccc}
        & & & \multicolumn{2}{c}{OIM}
              & \multicolumn{2}{c}{DIM}\\
        Graph & Edges & \(E^\star\)
              & \(E_{\mathrm{best}}\) & \(N_{\mathrm{GS}}\)
              & \(E_{\mathrm{best}}\) & \(N_{\mathrm{GS}}\)\\
        \hline
        \(G_1\) & 78 &  -56 &  -56 & 32 &  -56 & \textbf{42}\\
        \(G_2\) & 128 &  -72 &  -72 & \textbf{57} &  -72 & 35\\
        \(G_3\) & 406 & -122 & -122 & \textbf{80} & -122 & 73\\
        \(G_4\) & 436 & -132 & -132 & 14 & -132 & \textbf{76}\\
    \end{tabular}
    \end{ruledtabular}
\end{table}

\section*{Discussion}
\setcounter{subsection}{0}

\noindent The distinction between phase-difference and phase-sum coupling can be understood directly from the underlying phase symmetry of an oscillator network. In an autonomous network, a uniform shift of the time origin produces the transformation
\begin{equation}
    (\theta_i,\theta_j)
    \rightarrow
    (\theta_i+\alpha,\theta_j+\alpha),
    \label{eq:common_shift}
\end{equation}
without changing the physical state of the coupled system~\cite{nakao2016phase}. Any pairwise interaction potential must therefore be invariant under this common phase shift,
\begin{equation}
    E(\theta_i+\alpha,\theta_j+\alpha)
    =
    E(\theta_i,\theta_j),
    \qquad \forall\,\alpha .
\end{equation}
This invariance implies that the interaction can depend only on the relative phase,
\(
    E(\theta_i,\theta_j)
    =
    F(\theta_i-\theta_j),
\)
which explains why phase-difference coupling arises naturally in autonomous oscillator networks.

Phase-sum coupling does not possess this common-shift symmetry. For example,
\begin{equation}
    -\cos(\theta_i+\theta_j)
    \rightarrow
    -\cos(\theta_i+\theta_j+2\alpha),
\end{equation}
under the transformation in Eq.~\eqref{eq:common_shift}. A phase-sum interaction therefore requires an external phase reference against which the oscillator phases are defined. In the present architecture, this reference is supplied by the coherent second-harmonic pump. The resulting pairwise interaction is consequently defined relative to the pump phase,
\(
    E_{ij}^{(+)}
    =
    -KW_{ij}
    \cos(\theta_i+\theta_j-\psi),
\)
where $\psi$ denotes the phase of the $2\omega$ modulation associated with the coupling edge. The pump thereby breaks the continuous common-phase symmetry of the autonomous network and permits the phase-sum channel to contribute to the slow phase dynamics.

\section*{Conclusion}

\noindent Our work explores the idea of that the interaction law can serve as an additional design degree of freedom in oscillator-based computing, alongside the choice of oscillator platform and network architecture.

Phase-difference coupling arises naturally in passively coupled autonomous oscillator networks, but its gradient-flow characteristics can be ``fragile'' in nonlinear hardware because it places stringent constraints on the phase relationships between the oscillator waveform and its phase response. Phase-sum coupling provides a distinct route in which integrability follows from the structure of the interaction itself. In the context of Ising machines, this creates an alternative continuous embedding of the same discrete objective, with different energy landscapes and dynamical properties that can be advantageous for different problem instances.

Nevertheless, the present implementation achieves this functionality at the cost of additional hardware complexity, requiring multiple coupling paths and an external phase reference. Reducing this overhead, identifying device or circuit primitives that realize phase-sum coupling more directly, and developing architectures that can switch between or combine multiple interaction laws represent important directions for future work.

\vspace{12pt}
\noindent 
\textbf{Acknowledgments}--- This material is based upon work supported by ARO award W911NF-24-1-0228.
\vspace{6pt}

\noindent 
\textbf{Data availability}--- The data that support the findings of this study are available from the corresponding author upon reasonable request.\vspace{6pt}

\noindent 
\textbf{Author Contributions}---
A.H., S.B., and N.S. conceptualized the study. A.H., E.M.H.E, K.L. and S.B. performed the formal analysis and simulations. A.H., S.B., and N.S. wrote the manuscript. K.C. and N.S. supervised the study. All authors discussed the results and commented on the manuscript. A.H. and S.B. contributed equally. \vspace{6pt}

\noindent 
\textbf{Competing interests}---
The authors declare no competing interests.

\section*{REFERENCES}
\def\bibsection{}

\SupplementReferencesfalse
\ifSupplementReferences\else

\fi
\ifSupplementReferences

\fi

\CombinedBeginSupplement

\title{Supplemental Material\\[1.5ex]
Phase-Sum Coupling for Gradient-Flow Dynamics in Oscillator Ising Machines}

\author{Abir~Hasan\textsuperscript{1\#},
Sajib~Bain\textsuperscript{1\#},
E.M.~Hasantha~Ekanayake\textsuperscript{1}, Kyle~Lee\textsuperscript{2}, Kerem~Camsari\textsuperscript{2},
Nikhil~Shukla\textsuperscript{1*}}

\affiliation{%
\textnormal{\textsuperscript{1}}University of Virginia, Charlottesville, VA, USA\\
\textnormal{\textsuperscript{2}}\mbox{University~of~California,~Santa~Barbara,~Santa~Barbara,~CA,~USA}\\
\textnormal{\textsuperscript{*}Email: ns6pf@virginia.edu}\\
\textnormal{\textsuperscript{\#}Equal Contribution}
}

\makeatletter\begingroup\let\label\@gobble\maketitle\endgroup\makeatother
\onecolumngrid 

\section{Definition of Symbols}
The symbols used in this Supplementary Material are defined as follows:
\begin{itemize}
    \item $\theta_{i}$: Phase of the $i^{\text{th}}$ oscillator.
    \item $\Delta_{ij} = \theta_i - \theta_j$ (phase difference between oscillators $i$ and $j$), with $\Delta_{ji} = \theta_j - \theta_i = -\Delta_{ij}$.
    \item $E(\cdot)$: Scalar energy function.
    \item $Z(t)$: Infinitesimal phase response curve (iPRC).
    \item $s(t)$: Injected coupling waveform.
    \item $H(\Delta_{ij})$: Averaged pairwise interaction function.
    \item $H_{\mathrm{even}}(\Delta_{ij}), H_{\mathrm{odd}}(\Delta_{ij})$: Even and odd components of $H(\Delta_{ij})$, respectively.
    \item $\alpha_{n},\beta_{n}$: Amplitude of the $n^{\text{th}}$ harmonic of $Z(t)$ and $s(t)$, respectively.
    \item $\chi_{n}^{Z},\chi_{n}^{s}$: Phase of the $n^{\text{th}}$ harmonic of $Z(t)$ and $s(t)$, respectively.
    \item $\delta$: Measure of non-gradient behavior.
    \item $T$: Oscillation period; $\omega=2\pi/T$.
    \item $\tau = RC$: Time constant of each inverter stage in ring oscillator.
    \item $\varphi=\frac{(1+\sqrt{5})}{2}$ (Golden ratio).
    \item $H_{0}(\Delta_{ij})$: Base interaction function for the three-stage ring oscillator.
    \item $\omega$: Angular frequency. For our analysis, $\omega=1$.
    \item $\mathcal{C}$: Curl of the two-oscillator phase interaction over the phase space.
    \item $\mathcal{C}_{\text{RMS}}$:   Root mean square curl of the two-oscillator phase interaction over the phase space.
    \item $g_c(t)$: Trans-conductance of the modulated coupling element.
    \item $M$: Number of taps that produce uniformly spaced phase offsets of an oscillator waveform.
    \item $\psi_{ij}$: Reference phase of the signal used for modulating trans-conductance of the coupling element between oscillators $i$ and $j$.

\end{itemize}

\clearpage

\section{Supplementary Note 1}
\subsection{{Curl-free Condition for Gradient-Flow Dynamics}}

Here, we derive the necessary integrability condition under which the phase dynamics of a coupled oscillator network admit a scalar energy function. The existence of such a potential provides the basis for a gradient-flow interpretation of the phase dynamics. Its absence implies that the dynamics cannot, in general, be written as gradient flow on a scalar energy.\\

\noindent Consider a phase-reduced network of $N$ coupled oscillators,
\begin{align}
    \dot{\theta}_{i} = -H_{i}(\theta_{1},\ldots,\theta_{N}),
    \qquad i = 1,\ldots,N. \label{supp:eqn: phase_system}
\end{align}
We evaluate whether there exists a scalar function
$E(\theta_{1},\ldots,\theta_{N})$ such that
\begin{align}
    H_{i} = \dfrac{\partial E}{\partial \theta_{i}},
\end{align}
or equivalently, $\dot{\theta}_{i} = -\partial E/\partial \theta_{i}$. Under this condition, the dynamics correspond to gradient descent on the energy landscape given by $E$.\\

\noindent If such an energy function ${E}$ exists, Clairaut's theorem
implies
\begin{align}
    \dfrac{\partial^{2} E}{\partial \theta_{j}\,\partial \theta_{i}}
    =
    \dfrac{\partial^{2} E}{\partial \theta_{i}\,\partial \theta_{j}}. \label{supp:eq:Clairaut}
\end{align}
Combining Eq.~\eqref{supp:eq:Clairaut} with $H_{i} = \partial E/\partial \theta_{i}$, yields the \emph{integrability condition}
\begin{align}
    \dfrac{\partial H_{i}}{\partial \theta_{j}}
    =
    \dfrac{\partial H_{j}}{\partial \theta_{i}},
    \qquad \forall\, i\neq j.
    \label{supp:eq:integrability_condition}
\end{align}
Equation~\eqref{supp:eq:integrability_condition} implies that that the symmetry of the Jacobian matrix of the vector field is a necessary condition for the dynamics to admit a scalar potential.\\

\noindent\textbf{Corollary.} If for some pair $i \neq j$,
\begin{align}
    \dfrac{\partial H_{i}}{\partial \theta_{j}}
    \neq
    \dfrac{\partial H_{j}}{\partial \theta_{i}},
\end{align}
then no scalar function $E$ exists such that
$\dot{\theta} = -\nabla_{\theta_i} E$, and the dynamics carry a non-vanishing curl component.

\newpage
\section{Supplementary Note 2}
\subsection{Harmonic Quadrature Requirement for Gradient Flow}

\noindent 
Here, we derive the necessary and sufficient condition
under which the pairwise interaction function
\begin{align}
    H(\Delta_{ij})
    =
    \dfrac{1}{T}\int_{0}^{T} Z(t)\, s(t+\Delta_{ij})\,dt
    \label{supp:eq:H_def}
\end{align}
is odd i.e.,
$H(\Delta_{ji}) = -H(\Delta_{ij})$ upto an additive constant. 

\subsubsection{\textbf{Fourier Representation of the Interaction Function}}

Expanding the iPRC $Z(t)$ and the coupling waveform $s(t)$ in the Fourier form,
\begin{align}
    Z(t) &= \sum_{n}a_{n}\cos(nt)+b_{n}\sin(nt)
         = \sum_{n}\alpha_{n}\cos(nt-\chi_{n}^{Z}), \\
    s(t+\Delta_{ij}) &= \sum_{n}p_{n}\cos\!\big(n(t+\Delta_{ij})\big)
                  + q_{n}\sin\!\big(n(t+\Delta_{ij})\big) \nonumber \\
                &= \sum_{n}\beta_{n}\cos\!\big(n(t+\Delta_{ij})-\chi_{n}^{s}\big), 
\end{align}
where,
\begin{align}
    \alpha_{n} = \sqrt{a_{n}^{2}+b_{n}^{2}},\quad
    \beta_{n}  = \sqrt{p_{n}^{2}+q_{n}^{2}},\quad
\chi_n^Z = \tan^{-1}\!\left(\frac{b_n}{a_n}\right), \quad
\chi_n^s = \tan^{-1}\!\left(\frac{q_n}{p_n}\right).
\end{align}

\noindent Substituting into Eq.~\eqref{supp:eq:H_def} and using 
$\cos A \cos B = \tfrac{1}{2}[\cos(A+B)+\cos(A-B)]$, we obtain a double sum:
\begin{align}
H(\Delta_{ij})
&= \frac{1}{T}\int_0^{T}
   Z(t)\, s(t+\Delta_{ij})\,dt \\
&= \frac{1}{T}\int_0^{T}
   \sum_{n,m} \alpha_n \beta_m
   \cos(n t - \chi_n^Z)
   \cos(m(t+\Delta_{ij}) - \chi_m^s)\,dt 
   \qquad n,m \in \mathbb{N}.
\end{align}
Averaging over one period eliminates the cross terms with $n\neq m$ and the oscillatory terms at $2nt$, yielding:
\begin{align}
H(\Delta_{ij})
&= \frac{1}{T}\int_0^{T}
   \sum_n \alpha_n \beta_n
   \cos(n t - \chi_n^Z)
   \cos(n(t+\Delta_{ij}) - \chi_n^s)\,dt \\
&= \sum_n \frac{\alpha_n \beta_n}{2}
   \cos\!\left(n\Delta_{ij} + \chi_n^Z - \chi_n^s\right).
\label{supp:eq:H_fourier}
\end{align}

\subsubsection{\textbf{Decomposition of $H(.)$ into Even and Odd Components}}

Applying the angle-addition identity to
Eq.~\eqref{supp:eq:H_fourier},
\begin{align}
H(\Delta_{ij})
&= \sum_{n}\dfrac{\alpha_{n}\beta_{n}}{2}
   \Big[\cos(n\Delta_{ij})\cos(\chi_{n}^{Z}-\chi_{n}^{s})
        -\sin(n\Delta_{ij})\sin(\chi_{n}^{Z}-\chi_{n}^{s})\Big]
\nonumber \\
&= \underbrace{\sum_{n}\dfrac{\alpha_{n}\beta_{n}}{2}
   \cos(n\Delta_{ij})\cos(\chi_{n}^{Z}-\chi_{n}^{s})}_{H_{\mathrm{even}}(\Delta_{ij})}
   \;-\;
   \underbrace{\sum_{n}\dfrac{\alpha_{n}\beta_{n}}{2}
   \sin(n\Delta_{ij})\sin(\chi_{n}^{Z}-\chi_{n}^{s})}_{-H_{\mathrm{odd}}(\Delta_{ij})}.
\label{supp:eq: even_odd_decomp}
\end{align}

\subsubsection{\textbf{Condition for Oddness}}
Applying the integrability condition in Eq.~\eqref{supp:eq:integrability_condition} to the pairwise interaction function $H(\theta_i-\theta_j)$ yields
\begin{equation}
\frac{\partial H(\Delta_{ji})}{\partial \theta_j}
=
\frac{\partial H(\Delta_{ij})}{\partial \theta_i}
\;\;\Longrightarrow\;\;
H'(\Delta_{ji})=H'(\Delta_{ij}),
\end{equation}
This implies
\begin{equation}
H(\Delta_{ji})=-H(\Delta_{ij})+C,
\label{supp:eq: oddness}
\end{equation}
Thus, the interaction is odd up to an additive constant. Equivalently, $H(\Delta)-C/2$ must be odd. Any constant component of $H$ produces only a uniform phase drift and does not enter the integrability condition, which is determined by $H'$.

We therefore apply the oddness condition to the non-constant part of the interaction function. Using Eq.~\eqref{supp:eq:H_fourier}, we express
\[
H(\Delta)=H_{\rm DC}+
\sum_{n\ge 1}
\frac{\alpha_n\beta_n}{2}
\cos(n\Delta+\chi_n^Z-\chi_n^s),
\]
the additive constant is $C=2H_{\rm DC}$. The remaining even component is
\[
H_{\rm even}^{\rm nc}(\Delta)=
\sum_{n\ge 1}
\frac{\alpha_n\beta_n}{2}
\cos(n\Delta)\cos(\chi_n^Z-\chi_n^s).
\]
For the interaction to be odd up to the additive constant, this non-constant even component must vanish for all $\Delta$. Since the functions $\{\cos(n\Delta)\}_{n\ge 1}$ are linearly independent over one period, each coefficient must vanish independently, giving
\[
\cos(\chi_n^Z-\chi_n^s)=0,
\qquad \forall n\ge 1 \ \text{with}\ \alpha_n\beta_n\neq 0.
\]

Equivalently,
\begin{align}
\chi_n^Z-\chi_n^s
= \pm \frac{\pi}{2}\pmod{\pi},
\qquad \forall\, n \text{ with } \alpha_n\beta_n\neq 0 .
\label{supp:eq:quadrature_per_harmonic}
\end{align}
Thus, the interaction function is odd, up to an additive constant, if and only if each active harmonic of the injected waveform is in quadrature with the corresponding harmonic of the phase response function. This establishes the harmonic-by-harmonic quadrature condition for gradient-flow dynamics. For the systems considered here, the additive constant is zero because the iPRC has no DC component, and hence the averaged interaction has zero mean.

\subsubsection{\textbf{Normalized Measure of the Even Component}}

The oddness condition derived above can equivalently be expressed through the
symmetry of the derivative, $H'(\Delta_{ji})=H'(\Delta_{ij})$, where
$\Delta_{ji}=-\Delta_{ij}$. Differentiating Eq.~\eqref{supp:eq:H_fourier} gives
\begin{align}
H'(\Delta_{ij})
= -\sum_{n\geq1} \frac{n\alpha_n\beta_n}{2}
\sin\!\left(n\Delta_{ij}+\chi_n^Z-\chi_n^s\right).
\label{supp:eq:Hprime_fourier}
\end{align}
We express the deviation from this symmetry as 
\begin{align}
H'(\Delta_{ji})-H'(\Delta_{ij})
=
\sum_{n\geq1} \left[n\alpha_n\beta_n
\cos(\chi_n^Z-\chi_n^s)\right]\sin(n\Delta_{ij}).
\label{supp:eq:curl_expansion}
\end{align}
Equation~\eqref{supp:eq:curl_expansion} shows that the non-gradient contribution associated with the $n^{\text{th}}$ harmonic is weighted by $n\alpha_n\beta_n\cos(\chi_n^Z-\chi_n^s)$. \\

\noindent We therefore define the aggregate harmonic-weighted magnitude of the non-constant even component as
\begin{align}
\epsilon
=
\sum_{n\ge 1} n\alpha_n\beta_n
\left|\cos(\chi_n^Z-\chi_n^s)\right|.
\label{supp:eq:epsilon_def}
\end{align}
This quantity is zero only when every active harmonic satisfies the quadrature condition,
$\chi_n^Z-\chi_n^s=\pm \pi/2 \pmod{\pi}$, and becomes large when spectrally significant harmonics deviate from quadrature.\\

\noindent To obtain a dimensionless measure, we normalize $\epsilon$ by the maximum
possible harmonic-weighted even contribution,
\begin{align}
\epsilon_{\max}
=
\sum_{n\geq1} n\alpha_n\beta_n,
\label{supp:eq:norm}
\end{align}
which is attained when all active harmonics contribute coherently to the even
component. This yields 
\begin{align}
\delta
=
\sum_{n \geq 1} \delta_n
=
\frac{\epsilon}{\epsilon_{\max}}
=
\frac{
\sum_{n \geq 1} n\alpha_n\beta_n
|\cos(\chi_n^Z-\chi_n^s)|
}{
\sum_{m \geq 1} m\alpha_m\beta_m
},
\label{supp:eq:delta_def}
\end{align}
where the normalized harmonic-resolved contribution is
\begin{align}
\delta_n
=
\frac{
n\alpha_n\beta_n
|\cos(\chi_n^Z-\chi_n^s)|
}{
\sum_{m \geq 1} m\alpha_m\beta_m
}.
\end{align}
By construction, $\delta\in[0,1]$. The limiting case $\delta=1$
corresponds to a purely even interaction, while $\delta=0$ indicates a vanishing net even contribution. Strict gradient-flow behavior, however, requires the stronger harmonic-by-harmonic quadrature condition in Eq.~\eqref{supp:eq:quadrature_per_harmonic}. Intermediate values of $\delta$ quantify the extent to which harmonic misalignment induces deviations from energy-based dynamics.
\newpage

\section{Supplementary Note 3}
\subsection{Interaction Function for a Three-Stage Ring Oscillator}

\noindent In this section, we derive the iPRC for a  ring oscillator (3-stages) using the phase response formalism. Subsequently, we show that all tap-dependent interaction functions are shifted copies of a single base function $H_{0}$, and then evaluate $H_{0}$ in closed form. We note that the iPRC and the base interaction function have been derived in Refs.~\SupplementCite{supp:srivastava2007analytical,supp:bhansali2009gen}. We present them here for completeness and subsequently adapt them for the tap-dependent coupling. \\

\noindent We note that in electronic-oscillator literature, the infinitesimal phase response curve (iPRC) is commonly referred to as the perturbation projection vector~\SupplementCite{supp:demir1998phase}. With the oscillator frequency normalized to unity, the two are equivalent.

\begin{figure}[!h]
    \centering
    \includegraphics[scale=0.25]{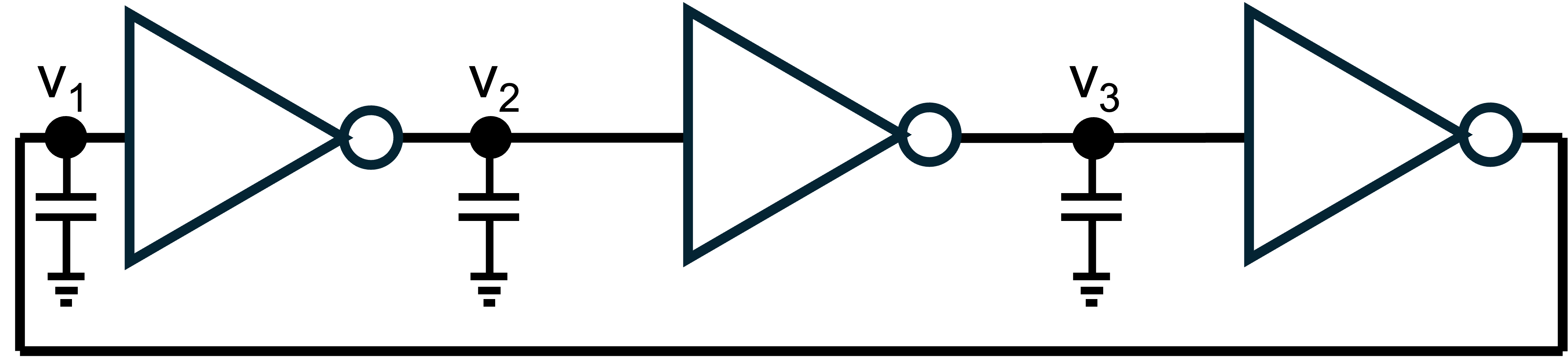}
    \caption{Schematic of a three-stage ring oscillator (RO). $v_1$, $v_2$, $v_3$, are the output taps.}
    \label{supp:fig:RO_3stage}
\end{figure}

\subsubsection{\textbf{Uncoupled Three-Stage Ring Oscillator}}

\noindent Consider the idealized three-stage ring oscillator model (Fig. \ref{supp:fig:RO_3stage}),
\begin{align}
    \dot{v}_{1} &= \dfrac{f(v_{3})-v_{1}}{\tau},\qquad
    \dot{v}_{2}  = \dfrac{f(v_{1})-v_{2}}{\tau},\qquad
    \dot{v}_{3}  = \dfrac{f(v_{2})-v_{3}}{\tau},
    \label{supp:eq: ring_model}
\end{align}
where $f(v) = \tanh(-kv)=-\tanh(kv)$ and $\tau=RC$. The variables $v_i$ denote centered, normalized inverter outputs, $v_i\in[-1,1]$, obtained by an affine rescaling of the physical CMOS voltage range $V_i\in[0,V_{\rm DD}]$. This normalization is used only for analytical convenience and does not affect the phase-reduced interaction symmetry or the metric $\delta$. \\

\noindent As derived in Refs.~\SupplementCite{supp:srivastava2007analytical,supp:bhansali2009gen}, the oscillator dynamics admit a periodic
steady-state solution of period
\begin{align}
    T = 6\ln(\varphi)\,\tau,\qquad
    \varphi = \dfrac{1+\sqrt{5}}{2},
    \label{supp:eq: ring_period}
\end{align}
and the three node voltages are shifted copies of a single scalar waveform $x(t)$,
\begin{align}
    v_{1}(t) = x(t),\qquad
    v_{2}(t) = x\!\left(t-\tfrac{2T}{3}\right),\qquad
    v_{3}(t) = x\!\left(t-\tfrac{T}{3}\right).
    \label{supp:eq: ring_shift}
\end{align}
Over one period,
\begin{align}
    x(t) =
    \begin{cases}
        1-\varphi\, e^{-t/\tau}, & 0\le t<T/2, \\[2pt]
        -1+\varphi\, e^{-(t-T/2)/\tau}, & T/2\le t<T,
    \end{cases}
    \label{supp:eq: x_waveform}
\end{align}
extended $T$-periodically.\\

\noindent The corresponding iPRC components inherit the same cyclic shift
structure,
\begin{align}
    Z_1(t) = p\!\left(t-\tfrac{2T}{3}\right),\qquad
    Z_2(t) = p\!\left(t-\tfrac{T}{3}\right),\qquad
    Z_3(t) = p(t),
    \label{supp:eq: ppv_shift}
\end{align}
where
\begin{align}
    p(t) =
    \begin{cases}
        \tau c_{1}\, e^{t/\tau}, & 0\le t<T/2, \\[2pt]
        \tau c_{2}\, e^{t/\tau}, & T/2\le t<T,
    \end{cases}
    \qquad
    c_{1} = \dfrac{1}{\sqrt{5}},\qquad
    c_{2} = \dfrac{2}{\sqrt{5}}-1,
    \label{supp:eq: p_waveform}
\end{align}
extended $T$-periodically.

\subsubsection{\textbf{Phase Reduction Under Weak Coupling}}

\noindent Consider two identical ring oscillators labeled $a$ and $b$. Suppose
node $m\in\{1,2,3\}$ of oscillator $b$ injects a weak additive
perturbation into node $r\in\{1,2,3\}$ of oscillator $a$. The
perturbation vector is
\begin{align}
    b_{p}^{(a)}(t) = \kappa\,e_{r}\,v_{m}^{(b)}(t),
    \label{supp:eq: bp}
\end{align}
where $e_{r}$ is the $r^{\text{th}}$ Cartesian unit vector and
$\kappa\ll 1$ represents a small coupling coefficient.\\

\noindent Using the iPRC phase model,
\begin{align}
    \dot{\alpha}_{a}(t) =
    Z^{\top}\!\big(t+\alpha_{a}(t)\big)\,b_{p}^{(a)}(t),
    \label{supp:eq: ppv_macro}
\end{align}
we obtain
\begin{align}
    \dot{\alpha}_{a}(t) =
    \kappa\,Z_{r}\!\big(t+\alpha_{a}(t)\big)\,v_{m}^{(b)}(t).
\end{align}
Averaging over one oscillation period yields the reduced phase
equation
\begin{align}
    \dot{\theta}_{a} = \omega + \kappa\, H_{r\leftarrow m}(\theta_{b}-\theta_{a}),
    \label{supp:eq: phase_eq}
\end{align}
where the tap-dependent interaction function is
\begin{align}
    H_{r\leftarrow m}(\Delta) =
    \dfrac{1}{T}\int_{0}^{T}\!
    Z_{r}(t)\,v_{m}(t+\Delta)\,dt.
    \label{supp:eq: Hrm_def}
\end{align}

\subsubsection{\textbf{Shift Symmetry of the Interaction Function}}

Since both the steady-state waveform and the iPRC inherit the cyclic symmetry of the ring, every tap-dependent interaction law is
a shifted copy of a single base interaction function.\\

\noindent Expressing $v_{m}(t) = x(t-\sigma_{m})$ and
$Z_{r}(t) = p(t-\rho_{r})$ with
\begin{align}
    (\sigma_{1},\sigma_{2},\sigma_{3}) &=
    \left(0,\,\tfrac{2T}{3},\,\tfrac{T}{3}\right),
    \qquad
    (\rho_{1},\rho_{2},\rho_{3}) =
    \left(\tfrac{2T}{3},\,\tfrac{T}{3},\,0\right),
    \label{supp:eq: shift_params}
\end{align}
and substituting into Eq.~\eqref{supp:eq: Hrm_def} gives
\begin{align}
    H_{r\leftarrow m}(\Delta) =
    \dfrac{1}{T}\int_{0}^{T}\!
    p(t-\rho_{r})\,x(t+\Delta-\sigma_{m})\,dt.
\end{align}
With the substitution $u=t-\rho_{r}$ and using $T$-periodicity, this
becomes
\begin{align}
    H_{r\leftarrow m}(\Delta) =
    \dfrac{1}{T}\int_{0}^{T}\!
    p(u)\,x\!\big(u+\Delta+\rho_{r}-\sigma_{m}\big)\,du.
\end{align}
Defining the \emph{base interaction function},
\begin{align}
    H_{0}(\Delta) \equiv
    \dfrac{1}{T}\int_{0}^{T}\! p(t)\,x(t+\Delta)\,dt,
    \label{supp:eq: H0_def}
\end{align}
we obtain
\begin{align}
    H_{r\leftarrow m}(\Delta) = H_{0}(\Delta + \beta_{rm}),
    \qquad
    \beta_{rm} = \rho_{r}-\sigma_{m}\!\!\pmod{T}.
    \label{supp:eq: H_shift_form}
\end{align}

\noindent All tap-dependent couplings thus share a single functional
form; changing the transmit or receive tap merely translates the
phase argument.

\subsubsection{\textbf{Closed-Form Evaluation of $H_{0}(\Delta)$}}

\noindent We now evaluate the base interaction function $H_0(\Delta)$ in closed form, following the analytical approach of Refs.~\SupplementCite{supp:srivastava2007analytical,supp:bhansali2009gen}. Since both the steady-state waveform $x(t)$ and the iPRC $p(t)$ are piecewise exponential with a breakpoint at $t=T/2$, the correlation integral in Eq.~\eqref{supp:eq: H0_def} must be evaluated piecewise in $\Delta$. We define
\begin{align}
    x_{1}(t) &= 1-\varphi\, e^{-t/\tau},
    &
    x_{2}(t) &= -1+\varphi\, e^{-(t-T/2)/\tau}, \\
    p_{1}(t) &= \tau c_{1}\, e^{t/\tau},
    &
    p_{2}(t) &= \tau c_{2}\, e^{t/\tau},
\end{align}
where $x(t)=x_1(t)$ and $p(t)=p_1(t)$ for $0\le t<T/2$, while
$x(t)=x_2(t)$ and $p(t)=p_2(t)$ for $T/2\le t<T$.

\begin{enumerate}
    \item[\textbf{(i)}] \textbf{Case 1: $0\le \Delta <T/2$.}

    In this interval, the shifted waveform $x(t+\Delta)$ changes branch at
    $t=T/2-\Delta$ and wraps around at $t=T-\Delta$. Thus,
    \begin{align}
        H_{0}(\Delta)
        =
        \frac{1}{T}
        \big[
        I_{1}(\Delta)+I_{2}(\Delta)+I_{3}(\Delta)+I_{4}(\Delta)
        \big],
        \label{supp:eq:H0_case1}
    \end{align}
    with
    \begin{align}
        I_{1}(\Delta) &= \int_{0}^{T/2-\Delta}
        p_{1}(t)\,x_{1}(t+\Delta)\,dt,\\
        I_{2}(\Delta) &= \int_{T/2-\Delta}^{T/2}
        p_{1}(t)\,x_{2}(t+\Delta)\,dt,\\
        I_{3}(\Delta) &= \int_{T/2}^{T-\Delta}
        p_{2}(t)\,x_{2}(t+\Delta)\,dt,\\
        I_{4}(\Delta) &= \int_{T-\Delta}^{T}
        p_{2}(t)\,x_{1}(t+\Delta-T)\,dt.
    \end{align}
    Direct evaluation yields
    \begin{widetext}
    \begin{align}
        I_{1}(\Delta)
        &= c_{1}\tau^{2}\!\left(e^{(T/2-\Delta)/\tau}-1\right)
        - c_{1}\tau\varphi\, e^{-\Delta/\tau}
        \left(\frac{T}{2}-\Delta\right),\\
        I_{2}(\Delta)
        &= -c_{1}\tau^{2}\!\left(e^{T/(2\tau)}-e^{(T/2-\Delta)/\tau}\right)
        + c_{1}\tau\varphi\, e^{(T/2-\Delta)/\tau}\Delta,\\
        I_{3}(\Delta)
        &= -c_{2}\tau^{2}\!\left(e^{(T-\Delta)/\tau}-e^{T/(2\tau)}\right)
        + c_{2}\tau\varphi\, e^{(T/2-\Delta)/\tau}
        \left(\frac{T}{2}-\Delta\right),\\
        I_{4}(\Delta)
        &= c_{2}\tau^{2}\!\left(e^{T/\tau}-e^{(T-\Delta)/\tau}\right)
        - c_{2}\tau\varphi\, e^{(T-\Delta)/\tau}\Delta.
    \end{align}
    \end{widetext}

    \item[\textbf{(ii)}] \textbf{Case 2: $T/2\le \Delta <T$.}

    In this interval, the wrap-around of $x(t+\Delta)$ occurs before the midpoint crossing. The base interaction function is therefore
    \begin{align}
        H_{0}(\Delta)
        =
        \frac{1}{T}
        \big[
        J_{1}(\Delta)+J_{2}(\Delta)+J_{3}(\Delta)+J_{4}(\Delta)
        \big],
        \label{supp:eq:H0_case2}
    \end{align}
    where
    \begin{align}
        J_{1}(\Delta) &= \int_{0}^{T-\Delta}
        p_{1}(t)\,x_{2}(t+\Delta)\,dt,\\
        J_{2}(\Delta) &= \int_{T-\Delta}^{T/2}
        p_{1}(t)\,x_{1}(t+\Delta-T)\,dt,\\
        J_{3}(\Delta) &= \int_{T/2}^{3T/2-\Delta}
        p_{2}(t)\,x_{1}(t+\Delta-T)\,dt,\\
        J_{4}(\Delta) &= \int_{3T/2-\Delta}^{T}
        p_{2}(t)\,x_{2}(t+\Delta-T)\,dt.
    \end{align}
    Evaluating these terms yields
    \begin{widetext}
    \begin{align}
        J_{1}(\Delta)
        &= -c_{1}\tau^{2}\!\left(e^{(T-\Delta)/\tau}-1\right)
        + c_{1}\tau\varphi\, e^{(T/2-\Delta)/\tau}(T-\Delta),\\
        J_{2}(\Delta)
        &= c_{1}\tau^{2}\!\left(e^{T/(2\tau)}-e^{(T-\Delta)/\tau}\right)
        - c_{1}\tau\varphi\, e^{(T-\Delta)/\tau}
        \left(\Delta-\frac{T}{2}\right),\\
        J_{3}(\Delta)
        &= c_{2}\tau^{2}\!\left(e^{(3T/2-\Delta)/\tau}-e^{T/(2\tau)}\right)
        - c_{2}\tau\varphi\, e^{(T-\Delta)/\tau}(T-\Delta),\\
        J_{4}(\Delta)
        &= -c_{2}\tau^{2}\!\left(e^{T/\tau}-e^{(3T/2-\Delta)/\tau}\right)
        + c_{2}\tau\varphi\, e^{(3T/2-\Delta)/\tau}
        \left(\Delta-\frac{T}{2}\right).
    \end{align}
    \end{widetext}
\end{enumerate}
Equations~\eqref{supp:eq:H0_case1} and~\eqref{supp:eq:H0_case2} provide the complete closed-form expression for $H_0(\Delta)$ over one period, $\Delta\in[0,T)$. Values outside this interval are obtained by $T$-periodic extension. Tap-dependent interaction functions then follow directly from the shift relation in Eq.~\eqref{supp:eq: H_shift_form}.

\newpage
\section{Supplementary Note 4}
\subsection{Ring Oscillator with a longer Inverter Chain $(N=25)$}

Here, we analyze the coupling characteristics of a longer-chain ring oscillator with $N=25$ inverter stages. The steady-state waveform $v(t)$ and the corresponding iPRC $Z(t)$ are computed numerically and decomposed into the first ten Fourier harmonics. Using these spectra, we evaluate the normalized harmonic-resolved contributions $\delta_n$ and their cumulative sum $\sum_n \delta_n$ to quantify the deviation from gradient-flow dynamics.

\begin{figure}[!h]
    \centering
    \includegraphics[width=1\linewidth]{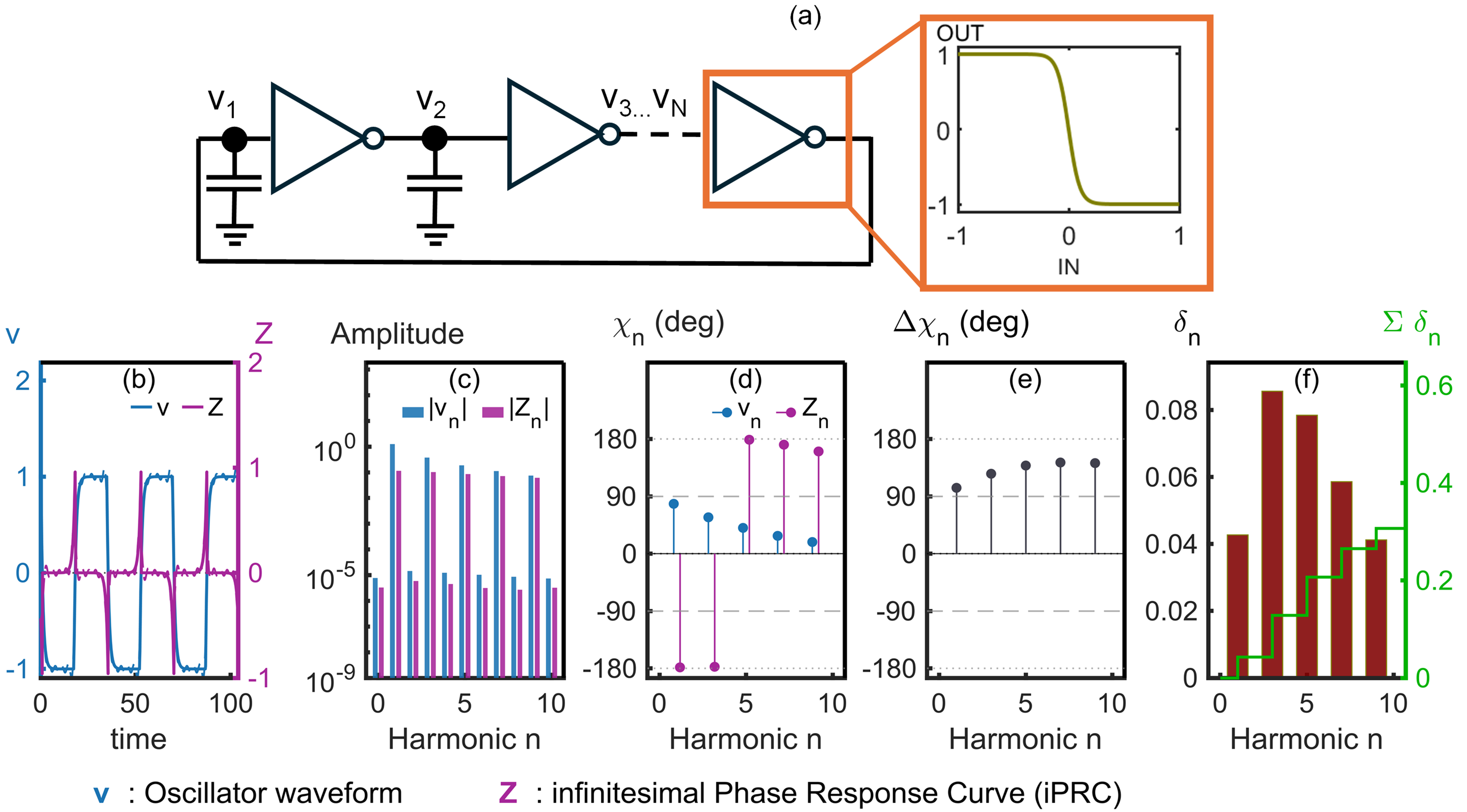}
    \caption{\justifying Harmonic analysis of a $N=25$ stage ring oscillator in a coupled two-oscillator system with coupling at the output taps. (a) Schematic of a three-stage ring oscillator. (b) Time-domain waveform $v(t)$ (left axis) and iPRC $Z(t)$ (right axis). (c) Amplitude; and (d) Phase spectra of the first ten harmonics of $v(t)$ and $Z(t)$, respectively. (e) Harmonic-wise phase difference, $\Delta\chi_n = \chi_n^Z - \chi_n^s$ (wrapped to $[-\pi,\pi]$), showing deviations from quadrature. (f) Normalized harmonic-resolved contributions $\delta_n$ to the even (non-gradient) component (left axis) and cumulative deviation $\delta$ (right axis).}
    \label{supp:fig: Fourier_Spectra}
\end{figure}

\newpage
\section{Supplementary Note 5}
\label{supp:supp:genearlized_phase_sum5}
\subsection{Candiate Energies for Dynamics with Even Interaction Components and their Time Deivatives} 

To examine the gradient-flow properties of oscillator networks
in the presence of both odd and even interaction components,
we derive the candidate energy functions and their time
derivatives for phase-sum and phase-difference dynamics.
This analysis provides the theoretical basis for comparing
the energy evolution of the two interaction laws.

Similar to the main text, we consider an interaction function containing first-harmonic odd and even components:
\begin{align}
    H(\theta_i+\theta_j)
    &=K_{o}\sin(\theta_i+\theta_j)+K_{e}\cos(\theta_i+\theta_j),
    \label{supp:eq:H_even_test_sum}\\
    H(\theta_i-\theta_j)
    &=K_{o}\sin(\theta_i-\theta_j)+K_{e}\cos(\theta_i-\theta_j),
    \label{supp:eq:H_even_test_diff}
\end{align}
for phase-sum and phase-difference coupling, respectively. Here, $K_{\mathrm{o}}$ and $K_{\mathrm{e}}$ are the coefficients of the odd and even components, respectively. With second-harmonic injection, relevant to the DIM and OIM,  the network dynamics are
\begin{align}
    \dot{\theta}_i^{(+)}
    &=-\sum_{j\ne i}W_{ij}H(\theta_i+\theta_j)
      -K_s\sin(2\theta_i),
    \label{supp:eq:dim_even_dynamics}\\
    \dot{\theta}_i^{(-)}
    &=-\sum_{j\ne i}W_{ij}H(\theta_i-\theta_j)
      -K_s\sin(2\theta_i),
    \label{supp:eq:oim_even_dynamics}
\end{align}
where $\dot{\theta}_i^{(+)}$ and $\dot{\theta}_i^{(-)}$ denote the phase-sum-based and phase-difference-based dynamics with the interaction function considered in Eq.~\eqref{supp:eq:H_even_test_sum}~and~\eqref{supp:eq:H_even_test_diff}. For $K_o=K$ and $K_e=0$, they reduce to the ideal models of Eqs.~2~and~1, respectively.

Extending the pairwise construction of 
\begin{equation}
    E_{ij}
    =
    W_{ij}
    \int^{\theta_i+\theta_j} H(\Theta)\,d\Theta,
    \label{supp:eq:general_phase_sum_potential}
\end{equation}
to a network with symmetric weights and $W_{ii}=0$, we define the following candidate energies, including the second-harmonic injection,
\begin{align}
    E_{+}
    &=\sum_{i,j}W_{ij}\left[-K_{\mathrm{o}}\cos(\theta_i+\theta_j)
      +K_{\mathrm{e}}\sin(\theta_i+\theta_j)\right]
    -K_s\sum_i\cos(2\theta_i),
    \label{supp:eq:dim_even_energy}\\
    E_{-}
    &=\sum_{i,j}W_{ij}\left[-K_{\mathrm{o}}\cos(\theta_i-\theta_j)
      +K_{\mathrm{e}}\sin(\theta_i-\theta_j)\right]
    -K_s\sum_i\cos(2\theta_i),
    \label{supp:eq:oim_even_energy}
\end{align}
for the phase-sum and phase-difference dynamics, respectively. We now determine whether their gradients reproduce the corresponding network dynamics through $\dot{\theta}_i^{(\pm)}=-\alpha\nabla_{\theta_i} E_{\pm}$, as required for these functions to serve as scalar potentials.

$E_+$ obeys $\dot{\theta}_i^{(+)}=-\tfrac{1}{2}\,\partial E_{+}/\partial\theta_i$. Therefore,
\begin{equation}
    \frac{dE_{+}}{dt}
    =\sum_i\frac{\partial E_{+}}{\partial\theta_i}\,\dot{\theta}_i^{(+)}
    =-2\sum_i\left(\dot{\theta}_i^{(+)}\right)^2
    \leq 0
    \label{supp:eq:dim_energy_rate}
\end{equation}
for any value of $K_e$. Thus, $E_{+}$ is a scalar potential of the phase-sum-based dynamics even in the presence of an even interaction component.

For the phase-difference-based dynamics, the $K_e\sin(\theta_i-\theta_j)$ contributions to $E_{-}$ (Eq.~\eqref{supp:eq:oim_even_energy}) cancel between the $(i,j)$ and $(j,i)$ terms for symmetric weights, whereas the even component remains in the phase dynamics. The phase velocity can be formulated as $\dot{\theta}_i^{(-)}=F_i^{\mathrm{o}}+F_i^{\mathrm{e}}$, where $F_i^{\mathrm{o}}=-\tfrac{1}{2}\,\partial E_{-}/\partial\theta_i$ contains the odd coupling and second-harmonic injection terms, and 
\begin{equation}
    F_i^{\mathrm{e}}
    =-K_e\sum_{j\ne i}W_{ij}\cos(\theta_i-\theta_j)
    \label{supp:eq:oim_even_force}
\end{equation}
is the contribution of the even component. The gradient of $E_{-}$ thus
reproduces only part of the phase-difference-based dynamics. Consequently,
\begin{equation}
    \frac{dE_{-}}{dt}
    =\sum_i\frac{\partial E_{-}}{\partial\theta_i}\,\dot{\theta}_i^{(-)}
    =-2\sum_i\left(F_i^{\mathrm{o}}\right)^2
     -2\sum_iF_i^{\mathrm{o}}F_i^{\mathrm{e}}.
    \label{supp:eq:oim_energy_rate}
\end{equation}
In Eq.~\eqref{supp:eq:oim_energy_rate}, the first term is non-positive,
whereas the second term can be positive. When the second term is positive and exceeds the magnitude of the first, $dE_{-}/dt>0$. Thus, $E_{-}$ is not guaranteed to decrease along phase-difference trajectories. Moreover, because its gradient omits the even interaction component, $E_{-}$ does not generate the full phase-difference dynamics when $K_{\mathrm{e}}\neq0$.

\newpage

\section{Supplementary Note 6}
\label{supp:supp:coupling_element}
\subsection{Transistor-Level Realization of the Coupling Element}
\noindent This note derives the port-current relation of the coupling circuit in Fig.~5(g) of the main text and establishes how the applied bias controls its effective coupling transconductance \(G_c\). The circuit produces differential output currents in response to the voltage difference between ports \(A\) and \(B\).The diode-connected PMOS transistor $M_1$ carries the applied bias current $I_{\mathrm{bias}}(t)$, which is mirrored by $M_2$ to the common source of the matched PMOS differential pair $M_3$--$M_4$. The gates of $M_3$ and $M_4$ form coupling ports $A$ and $B$, respectively. The corresponding branch currents are replicated through the NMOS mirrors $M_5$--$M_8$ and PMOS mirrors $M_9$--$M_{12}$ and returned to the two coupling ports. Let
\begin{equation}
    \begin{aligned}
    m_T&\equiv\frac{(W/L)_2}{(W/L)_1},\\
    m_N&\equiv\frac{(W/L)_7}{(W/L)_5}
          =\frac{(W/L)_8}{(W/L)_6},\\
    m_P&\equiv\frac{(W/L)_{10}}{(W/L)_9}
          =\frac{(W/L)_{12}}{(W/L)_{11}}.
    \end{aligned}
    \label{supp:eq:coupler_mirror_ratios}
    \end{equation}
and let $m_R\equiv m_Nm_P$ be the gain of either matched return path. The implemented sizing uses $m_T=1$, $m_N=1$, and $m_P=2$, and hence $m_R=2$. Neglecting channel-length modulation, mirror mismatch, and gate current, the tail and returned currents satisfy~\SupplementCite{supp:gray2009analysis}
\begin{equation}
    \begin{aligned}
    I_T(t)&=m_T I_{\mathrm{bias}}(t),\\
    i_A(t)&=m_R I_{D3}(t),
    &i_B(t)&=m_R I_{D4}(t).
    \end{aligned}
    \label{supp:eq:coupler_mirror_chain}
\end{equation}
Let $v_d\equiv v_A-v_B$. Raising $v_A$ reduces $|V_{SG3}|$, so the PMOS pair steers current away from port $A$ and toward port $B$. At the balanced point, $v_d=0$, each input device carries $I_T(t)/2$. Let $g_m(t)$ denote the transconductance magnitude of either input device at that operating current. To first order in $v_d$,
\begin{equation}
    \begin{aligned}
    I_{D3}(t)&\simeq\frac{I_T(t)}{2}-\frac{g_m(t)}{2}v_d,\\
    I_{D4}(t)&\simeq\frac{I_T(t)}{2}+\frac{g_m(t)}{2}v_d.
    \end{aligned}
    \label{supp:eq:coupler_branch_currents}
\end{equation}
Substituting Eq.~\eqref{supp:eq:coupler_mirror_chain} gives the first-order differential-mode port relation,
\begin{equation}
    \begin{aligned}
    i_A(t)&\simeq I_{\mathrm{cm}}(t)-g_c(t)(v_A-v_B),\\
    i_B(t)&\simeq I_{\mathrm{cm}}(t)+g_c(t)(v_A-v_B).
    \end{aligned}
    \label{supp:eq:coupler_ports_linear}
\end{equation}
where
\begin{equation}
    \begin{aligned}
    I_{\mathrm{cm}}(t)&=\frac{m_Rm_T}{2}I_{\mathrm{bias}}(t),\\
    g_c(t)&=\frac{m_R}{2}
    g_m\!\left(\frac{m_T I_{\mathrm{bias}}(t)}{2}\right).
    \end{aligned}
    \label{supp:eq:coupler_Icm_Gc}
\end{equation}
The argument of $g_m$ in Eq.~\eqref{supp:eq:coupler_Icm_Gc} denotes the instantaneous drain current of either input transistor. For the implemented ratios, $I_{\mathrm{cm}}(t)=I_{\mathrm{bias}}(t)$ and $g_c(t)=g_m(t)$. The differential returned-current transconductance is $2g_c(t)$, whereas $g_c(t)$ is the coefficient presented at either port and is therefore the quantity used in the phase-reduced coupling law.

\noindent For a long-channel device in strong inversion~\SupplementCite{supp:gray2009analysis}, let
\begin{equation}
    \beta_p\equiv\mu_p C_{\mathrm{ox}}(W/L)_{3,4},
    \qquad
    I_D=\frac{\beta_p}{2}V_{\mathrm{ov}}^2.
    \label{supp:eq:coupler_square_law}
    \end{equation}
    At the balanced operating point,
    \begin{equation}
     g_m(t)=\sqrt{\beta_p m_T I_{\mathrm{bias}}(t)},
    \end{equation}
    and Eq.~\eqref{supp:eq:coupler_Icm_Gc} becomes
    \begin{equation}
     g_c(t)=\frac{m_R}{2}
     \sqrt{\beta_p m_T I_{\mathrm{bias}}(t)}.
    \label{supp:eq:coupler_Gc_strong}
    \end{equation}
    For $m_T=1$ and $m_R=2$,
    \begin{equation}
     g_c(t)=
     \sqrt{\mu_p C_{\mathrm{ox}}(W/L)_{3,4}I_{\mathrm{bias}}(t)}.
    \label{supp:eq:coupler_Gc_implemented}
\end{equation}

\noindent In weak inversion, with slope factor $n$ and thermal voltage $V_T=kT/q$,

\begin{equation}
 g_m(t)=\frac{m_T I_{\mathrm{bias}}(t)}{2nV_T},
 \qquad
 g_c(t)=\frac{m_Rm_T}{4nV_T}I_{\mathrm{bias}}(t).
\label{supp:eq:coupler_Gc_weak}
\end{equation}

\noindent In the ideal weak-inversion model~\SupplementCite{supp:enz1995analytical}, \(g_c\) is proportional to \(I_{\mathrm{bias}}\) in Eq.~\eqref{supp:eq:coupler_Gc_weak}, so sinusoidal bias modulation at \(2\omega\) produces only the dc and \(2\omega\) components of \(g_c(t)\). In the ideal strong-inversion model, the square-root dependence on \(I_{\mathrm{bias}}\) in Eq.~\eqref{supp:eq:coupler_Gc_strong} generates additional harmonics at integer multiples of \(2\omega\).

\newpage

\section{Supplementary Note 7}
\subsection{Suppression of Cross-Harmonic Interactions through Phase-Resolved Coupling}

\begin{figure}[!h]
    \centering
    \includegraphics[width=\linewidth]{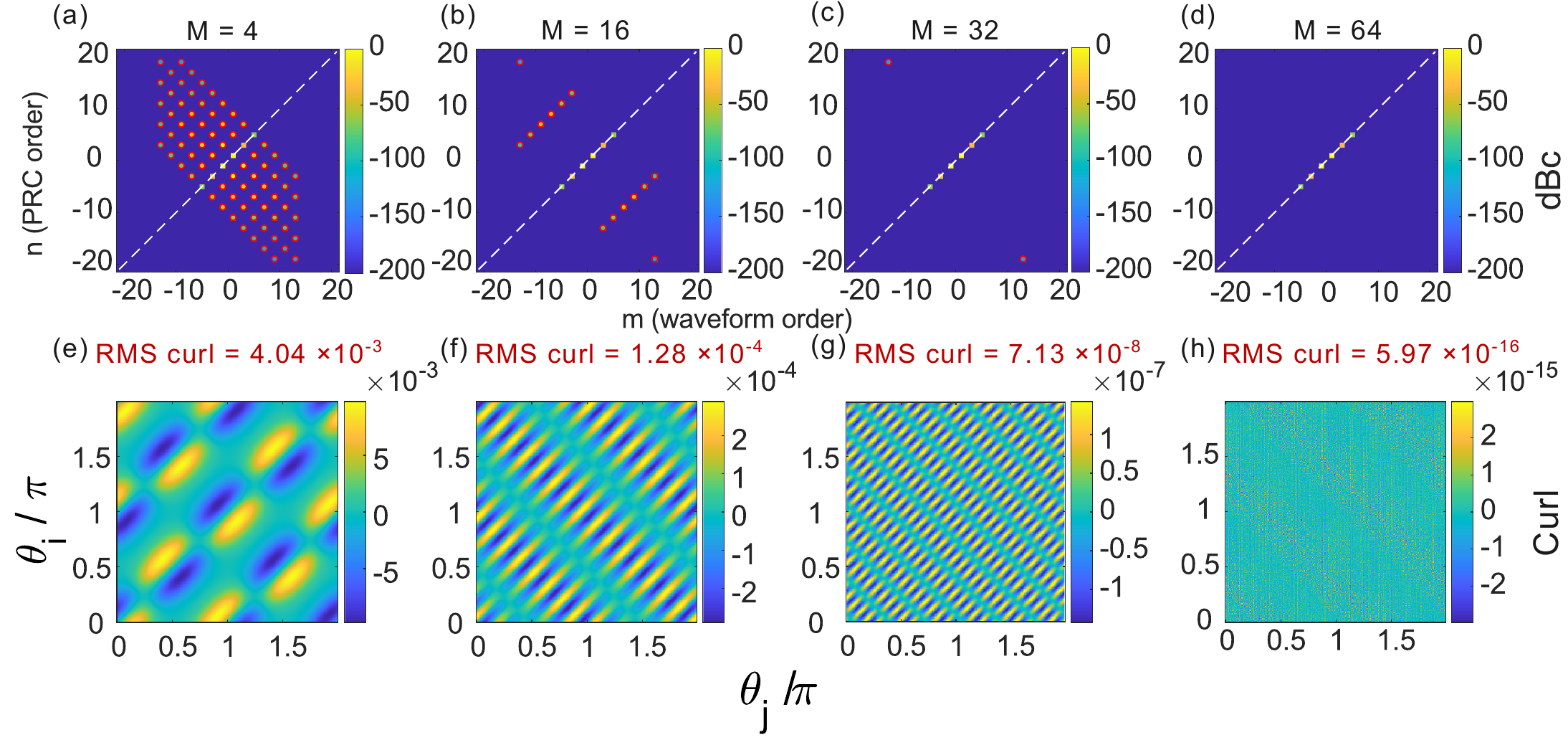}
    \caption{\justifying \small
    \textbf{Suppression of mixed-harmonic interactions with increasing coupling paths.}
    \textbf{(a--d)} Two-dimensional Fourier spectra of the averaged interaction for (M=4,16,32,) and (64), respectively. The dashed lines indicate the desired matched harmonics ((n=m)).
    \textbf{(e-h)} Corresponding curl of the phase dynamics, showing a progressive reduction in non-gradient contributions as (M) increases. The RMS curl is indicated in each panel.   
    }
\label{supp:fig:phase_resolved}
\end{figure}

In this section, we investigate how increasing the number of phase-resolved coupling paths progressively suppresses phase-domain aliasing and the resulting non-gradient contributions to the oscillator dynamics. To evaluate this effect, we consider a state-space model of a 32-stage differential ring oscillator, providing access to 64 phase-shifted taps. This allows us to systematically investigate the influence of the number of coupling paths, $M$, on the resulting phase interactions and their gradient-flow characteristics.

To quantify deviations from gradient flow, we evaluate the curl of the two-oscillator phase interaction field and its root-mean-square (RMS) value over the phase space, defined as

\begin{equation}
\begin{aligned}
\mathcal{C}(\theta_1,\theta_2)
&=
\frac{\partial H_1}{\partial \theta_2}
-
\frac{\partial H_2}{\partial \theta_1},
\\[6pt]
\mathcal{C}_{\mathrm{RMS}}
&=
\sqrt{
\frac{1}{(2\pi)^2}
\int_0^{2\pi}\int_0^{2\pi}
\mathcal{C}^2(\theta_1,\theta_2)\,
d\theta_1\,d\theta_2
}.
\end{aligned}
\end{equation}

Here, $H_1$ and $H_2$ denote the phase interaction functions of the two coupled oscillators. The RMS curl provides a quantitative measure of the non-gradient contributions to the phase dynamics, with $\mathcal{C}_{\mathrm{RMS}}=0$ corresponding to a curl-free interaction field.

To construct the phase-sum interactions, we extract the harmonic spectra of the oscillator waveform and its infinitesimal phase response curve (iPRC), retaining components within $60$ dB of their respective fundamental harmonics ($\geq -60$ dBc). The coupling modulation is represented by its DC component and ten appreciable harmonic components. The resulting interaction is evaluated using the dominant contributions from the first 32 harmonic orders.

We first consider $M=4$ phase-resolved coupling paths. As shown in Fig.~\ref{supp:fig:phase_resolved}(a), the two-dimensional harmonic spectrum contains significant off-diagonal components ($n\neq m$), corresponding to cross-harmonic interactions that survive phase-domain sampling. These aliased contributions introduce phase dependences beyond the desired phase-sum interaction and generate non-gradient components in the reduced phase dynamics. The corresponding curl distribution in the $(\theta_1,\theta_2)$ phase space is shown in Fig.~\ref{supp:fig:phase_resolved}(e), yielding an RMS curl of approximately $4.04\times10^{-3}$.

As the number of coupling paths increases to $M=16$, $32$, and $64$, the residual off-diagonal harmonic contributions are progressively suppressed, as illustrated in Figs.~\ref{supp:fig:phase_resolved}(b)--(d). This suppression is accompanied by a substantial reduction in the non-gradient curl, as shown in Figs.~\ref{supp:fig:phase_resolved}(f)--(h). Specifically, the RMS curl decreases to $1.28\times10^{-4}$ for $M=16$ and $7.13\times10^{-8}$ for $M=32$, before reaching $<10^{-15}$ for $M=64$, effectively vanishing to numerical precision.

These results are consistent with the Nyquist-like phase-domain sampling criterion derived in Section D of the main manuscript. Increasing the number of phase-resolved coupling paths suppresses an increasingly broad range of unwanted cross-harmonic interactions while preserving the matched-harmonic contributions responsible for phase-sum coupling. Consequently, sufficiently dense phase-domain sampling recovers the intrinsic gradient-flow structure of phase-sum interactions within the considered harmonic bandwidth.

\newpage

\section{Supplementary Note 8}
\subsection{Fixed Point and Stability Analysis}
\label{supp:supp:fixed_point_analysis} 
On the binary manifold, phase-difference and phase-sum coupling encode the same Ising objective, so the binary state of a network does not identify the coupling. The nonbinary fixed points prior to SHI injection provide the diagnostic. Here, we derive these fixed points and their linear stability for the two- and three-oscillator networks considered in the main text. We assume harmonic oscillators with \(K>0\) and AFM coupling, \(W_{ij}=-1\) on each edge, with all edges sharing the pump phase \(\psi\).

\subsection{Two-oscillator network}

\textit{Phase-difference coupling.}
The reduced dynamics are
\begin{equation}
\begin{aligned}
    \dot{\theta}_1
    &=K\sin(\theta_1-\theta_2),\\
    \dot{\theta}_2
    &=K\sin(\theta_2-\theta_1).
\end{aligned}
\label{supp:eq:fp_two_oim}
\end{equation}
The opposite-phase configuration
\begin{equation}
    \theta_1^*=\theta_0,
    \qquad
    \theta_2^*=\theta_0+\pi,
    \label{supp:eq:fp_two_oim_opposite_state}
\end{equation}
is a fixed point for any common phase offset \(\theta_0\). Linearization about this fixed point gives
\begin{equation}
    \lambda_{\mathrm{OIM}}^{(2)}
    =
    \{0,-2K\}.
    \label{supp:eq:fp_two_oim_eigenvalues}
\end{equation}
The zero eigenvalue corresponds to a common phase shift, while the negative eigenvalue stabilizes the phase difference at \(\pi\).

\textit{Phase-sum coupling.}
The reduced dynamics are
\begin{equation}
    \dot{\theta}_1
    =
    \dot{\theta}_2
    =
    K\sin(\theta_1+\theta_2-\psi_{12}).
    \label{supp:eq:fp_two_dim}
\end{equation}
The fixed points with
\begin{equation}
    \theta_1^*+\theta_2^*-\psi_{12}=\pi
    \pmod{2\pi}
    \label{supp:eq:fp_two_dim_stable_sum}
\end{equation}
have
\begin{equation}
    \lambda_{\mathrm{DIM}}^{(2)}
    =
    \{-2K,0\}.
    \label{supp:eq:fp_two_dim_eigenvalues}
\end{equation}
The negative eigenvalue stabilizes the phase sum, while the zero eigenvalue corresponds to the phase difference, which remains at its initial value because the two phase velocities are equal. For equal initial phases, the pair settles at the common phase
\begin{equation}
    \theta_1^{*}
    =
    \theta_2^{*}
    =
    \frac{\psi_{12}+\pi}{2},
    \label{supp:eq:fp_two_dim_common}
\end{equation}
or at its counterpart with both phases shifted by \(\pi\). For \(\psi_{12}=0\), the common phase is \(\pi/2\).

\subsection{Three-oscillator network}

In the fully connected three-oscillator network, the neutral relative-phase direction of the phase-sum-coupled pair is removed.

\textit{Phase-difference coupling.}
The reduced dynamics are
\begin{equation}
    \dot{\theta}_i
    =
    K\sum_{j\neq i}\sin(\theta_i-\theta_j),
    \qquad i\in\{1,2,3\}.
    \label{supp:eq:fp_three_oim}
\end{equation}
A phase difference of \(\pi\) cannot hold on all three edges, so the network is frustrated. The equally spaced configuration
\begin{equation}
    \theta_i^*
    =
    \theta_0+\frac{2\pi}{3}(i-1),
    \qquad i\in\{1,2,3\},
    \label{supp:eq:fp_three_oim_state}
\end{equation}
is a fixed point for any common phase offset \(\theta_0\), since the two sine contributions cancel at each oscillator. Linearization about this fixed point gives
\begin{equation}
    \lambda_{\mathrm{OIM}}^{(3)}
    =
    \left\{
        0,-\frac{3K}{2},-\frac{3K}{2}
    \right\}.
    \label{supp:eq:oim_120_eigenvalues}
\end{equation}
The zero eigenvalue corresponds to a common phase shift, while the two negative eigenvalues stabilize the \(2\pi/3\) phase separation.

\textit{Phase-sum coupling.}
The reduced dynamics are
\begin{equation}
    \dot{\theta}_i
    =
    K\sum_{j\neq i}
    \sin(\theta_i+\theta_j-\psi),
    \qquad i\in\{1,2,3\}.
    \label{supp:eq:three_oscillator_dim}
\end{equation}
The condition \(\theta_i+\theta_j-\psi=\pi\) modulo \(2\pi\) holds on all three edges at the common-phase fixed point
\begin{equation}
    \theta_1^{*}
    =
    \theta_2^{*}
    =
    \theta_3^{*}
    =
    \frac{\psi+\pi}{2},
    \label{supp:eq:dim_in_phase_state}
\end{equation}
and at its counterpart with all three phases shifted by \(\pi\). With \(\cos(\theta_i^*+\theta_j^*-\psi)=-1\) on every edge, linearization about this fixed point gives
\begin{equation}
    \lambda_{\mathrm{DIM}}^{(3)}
    =
    \{-4K,-K,-K\}.
    \label{supp:eq:dim_in_phase_eigenvalues}
\end{equation}
All three eigenvalues are negative, so the fixed point is stable. For \(\psi=0\), the common phase is \(\pi/2\), and the three oscillator waveforms are mutually in phase.

\newpage
\def\bibsection{\section*{References for Supplemental Material}}
\SupplementReferencestrue

\end{document}